\documentstyle[aps,eqsecnum,epsf,12pt]{revtex}

\newcommand{\bq}{\begin{quotation}\noindent}
\newcommand{\eq}{\end{quotation}}
\newcommand{\be}{\begin{equation}}
\newcommand{\ee}{\end{equation}}
\newcommand{\bea}{\begin{eqnarray}}
\newcommand{\eea}{\end{eqnarray}}
\newcommand{\bc}{\begin{center}}
\newcommand{\ec}{\end{center}}

\def\tr{{\rm tr}}

\def\vec#1{{\bf #1}}
\def\HD{{\cal H}_D}
\def\H{{\cal H}}
\def\ket#1{|#1\rangle}
\def\bra#1{\langle#1|}

\begin{document}

\draft

\title{Unknown Quantum States and Operations, a Bayesian View}

\author{Christopher A. Fuchs$^1$ and
R\"udiger Schack$^2$\medskip}

\address{
$^1$Quantum Information and Optics Research, Bell Labs, Lucent
Technologies,
\\
600--700 Mountain Avenue, Murray Hill, New Jersey 07974, USA
\\
$^3$Department of Mathematics, Royal Holloway, University of London,
\\
Egham, Surrey TW20$\;$0EX, UK}

\date{26 February 2004}

\maketitle

\begin{abstract}
The classical de Finetti theorem provides an operational definition
of the concept of an unknown probability in Bayesian probability
theory, where probabilities are taken to be degrees of belief instead
of objective states of nature.  In this paper, we motivate and review
two results that generalize de Finetti's theorem to the quantum
mechanical setting:  Namely a de Finetti theorem for quantum states
and a de Finetti theorem for quantum operations. The quantum-state
theorem, in a closely analogous fashion to the original de Finetti
theorem, deals with exchangeable density-operator assignments and
provides an operational definition of the concept of an ``unknown
quantum state'' in quantum-state tomography.  Similarly, the
quantum-operation theorem gives an operational definition of an
``unknown quantum operation'' in quantum-process tomography. These
results are especially important for a Bayesian interpretation of
quantum mechanics, where quantum states and (at least some) quantum
operations are taken to be states of belief rather than states of
nature.
\end{abstract}

\section{Introduction}
\label{sec-intro}

What is a quantum state?\,\footnote{This paper represents
predominantly a culling of the material in
Refs.~\protect\cite{Caves2002b,Schack2000,Fuchs2003b}. Everything,
however, has been updated to accommodate the major shift in our
thinking represented in
Refs.~\protect\cite{FuchsWHAT,Caves2002c,Fuchs2002}. In particular,
it reflects a change in our views of quantum probabilities from that
of an objective Bayesianism of the type promoted by E.~T.
Jaynes~\cite{Jaynes2003} to a subjective or personalistic Bayesianism
of the type promoted by B.~de Finetti, L.~J. Savage, J.~M. Bernardo
and A.~F.~M. Smith, and
R.~Jeffrey~\cite{DeFinetti1989,DeFinetti1990,Savage1972,Bernardo1994,%
Jeffrey2003}. This makes all the difference in the world with regard
to the meaning of quantum states, operations, and their usage within
statistical theory.} Since the earliest days of quantum theory, it
has been understood that the quantum state can be used (through the
Born rule) to derive probability distributions for the outcomes of
all measurements that can be performed on a quantum system. But is it
more than that? Is a quantum state an actual property of the system
it describes? The Bayesian view of quantum states
\cite{Caves2002b,Schack2000,Fuchs2003b,FuchsWHAT,Caves2002c,%
Fuchs2002,Fuchs2000,Caves2002a,Caves1996,Brun2001a,Fuchs2003a,%
Schack2003,Appleby2003,vanEnk2001,FuchsVaxjo,FuchsPaulian} is that it
is not: The quantum state is not something the system itself
possesses. Rather it is solely a function of the observer (or,
better, agent) who contemplates the predictions, gambles, decisions,
or actions he might make with regard to those quantum measurements.

What distinguishes this view from a more traditional
``Copenhagen-interpretation style'' view---for instance the view
expressed so clearly and carefully in Ref.~\cite{Peres1984}---is that
there is no pretense that a quantum state represents a physical fact.
Quantum states come logically before that: They represent the
temporary and provisional {\it beliefs\/} a physicist holds as he
travels down the road of inquiry. It is the outcomes of quantum
measurements that represent physical facts within quantum theory, not
the quantum states. In particular, there is no fact of nature to
prohibit two different agents from using distinct pure states
$|\psi\rangle$ and $|\phi\rangle$ for a single quantum
system.\,\footnote{Contrast this to the treatment of
Refs.~\cite{Mermin2001,Brun2001b,Mermin2002}.  In any case, the
present point does not imply that a single agent can believe
willy-nilly anything he wishes. To quote D.~M. Appleby, ``You know,
it is {\it really\/} hard to believe something you don't actually
believe.''} Difficult though this may be to accept for someone
trained in the traditional presentation of quantum mechanics, the
only thing it demonstrates is a careful distinction between the terms
{\it belief\/} and {\it fact}.

Quantum states are not facts.\,\footnote{For a selection of papers
that we believe help shore up this statement in various ways---though
most of them are not explicitly Bayesian in their view of
probability---see  Ref.~\cite{SugarLump}.} But if so, then what is an
``unknown quantum state''? There is hardly a paper in the field of
quantum information that does not make use of the phrase. Unknown
quantum states are teleported~\cite{Bennett1993,Experiments1998},
protected with quantum error correcting
codes~\cite{Shor1995,Steane1996}, and used to check for quantum
eavesdropping~\cite{Bennett1984,CryptoExperiments}.  The list of uses
grows each day. Are all these papers nonsense? In a Bayesian view of
quantum states, the phrase is an oxymoron, a contradiction in terms:
If quantum states are states of belief rather than states of nature,
then a state is {\it known\/} by someone---at the very least, by the
agent who holds it.

Thus for a quantum Bayesian, if a phenomenon ostensibly invokes the
concept of an unknown state in its formulation, the unknown state
must be a kind of shorthand for a more involved story.  In other
words, the usage should be viewed a call to arms, an opportunity for
further analysis. For any phenomenon using the idea of an unknown
quantum state, the quantum Bayesian should demand that either:
\begin{enumerate}
\item
The owner of the unknown state---some further agent---be explicitly
identified.  (In this case, the unknown state is merely a stand-in
for the unknown state of belief of an essential player who went
unrecognized in the original formulation.) Or,
\item
If there is clearly no further agent upon the scene, then a way must
be found to reexpress the phenomenon with the term ``unknown state''
banished from the formulation. (In this case, the end-product will be
a single quantum state used for describing the phenomenon---namely,
the state that actually captures the initial agent's overall beliefs
throughout.)
\end{enumerate}

In this paper, we will analyze the particular use of unknown states
that comes from {\it quantum-state
tomography\/}~\cite{Vogel1989b,Smithey1993,Leonhardt1995}. Beyond
that, we will also argue for the necessity of (and carry out) a
similar analysis for {\it quantum-process
tomography\/}~\cite{Turchette1995b,Chuang1997,Poyatos1997}.

The usual, non-Bayesian description of quantum-state tomography is
this. A device of some sort repeatedly prepares many instances of a
quantum system in a fixed quantum state $\rho$, pure or mixed.  An
experimentalist who wishes to characterize the operation of the
device or to calibrate it for future use may be able to perform
measurements on the systems it prepares even if he cannot get at the
device itself. This can be useful if the experimenter has some prior
knowledge of the device's operation that can be translated into a
probability distribution over states. Then learning about the state
will also be learning about the device. Most importantly, though,
this description of tomography assumes the state $\rho$ is unknown.
The goal of the experimenter is to perform enough measurements, and
enough kinds of measurements (on a large enough sample), to estimate
the identity of $\rho$.

\begin{center}
\begin{figure} 
\epsfxsize=8cm \epsfbox{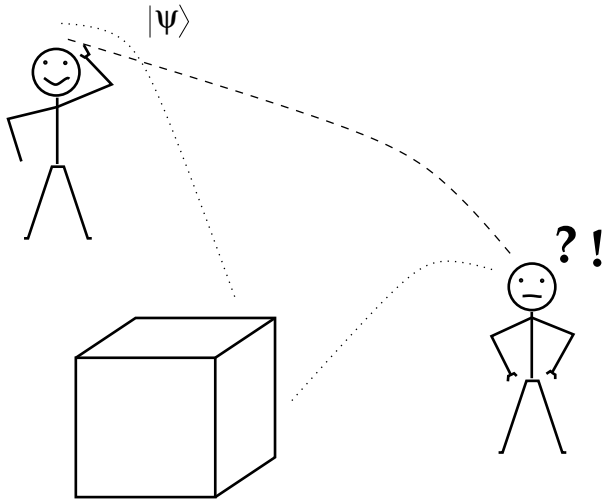} \bigskip\caption{What can the term
``unknown state'' mean if quantum states are taken to be solely
compendia of Bayesian expectations rather than states of nature? When
we say that a system has an unknown state, must we always imagine a
further agent whose state of belief is symbolized by some
$|\psi\rangle$, and it is the identity of that belief which we are
ignorant of?}
\end{figure}
\end{center}

This is clearly an example where there is no further player on whom
to pin the unknown state as a state of belief or judgment.  Any
attempt to find such a missing player would be entirely artificial:
Where would the player be placed?  On the inside of the device the
tomographer is trying to characterize? The only available course is
the second strategy above---to banish the idea of the unknown state
from the formulation of tomography.

\begin{center}
\begin{figure} \leavevmode
\epsfxsize=9cm \epsfbox{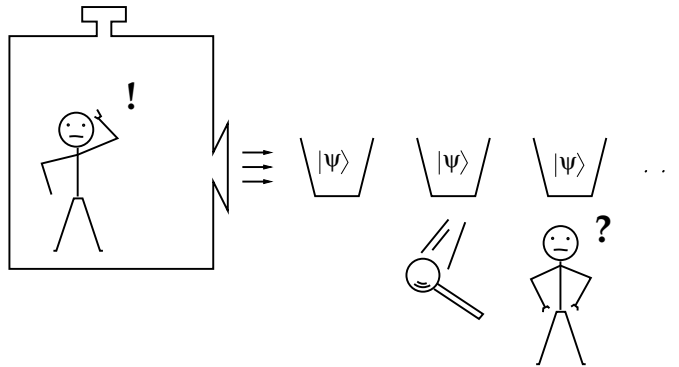} \bigskip\caption{To make sense of
quantum tomography, must we resort to imagining a ``man in the box''
who has a better description of the systems than we do? How contrived
would the Bayesian story be if this were so!}
\end{figure}
\end{center}

To do this, we take a cue from the field of Bayesian probability
theory
itself~\cite{DeFinetti1989,DeFinetti1990,Savage1972,Bernardo1994,%
Kyburg1980,DeFinettiCollected}. In Bayesian theory, probabilities are
not objective states of nature, but measures of personalistic belief.
The overarching Bayesian theme is to identify the conditions under
which a set of decision-making agents can come to a common belief or
probability assignment for a random variable even though their
initial beliefs differ.  Bernardo and Smith~\cite{Bernardo1994} make
the point forcefully:
\bq
\indent
[I]ndividual degrees of belief, expressed as probabilities, are
inescapably the starting point for descriptions of uncertainty. There
can be no theories without theoreticians; no learning without
learners; in general, no science without scientists.  It follows that
learning processes, whatever their particular concerns and fashions
at any given point in time, are necessarily reasoning processes which
take place in the minds of individuals.  To be sure, the object of
attention and interest may well be an assumed external, objective
reality:  but the actuality of the learning process consists in the
evolution of individual, subjective beliefs about that reality.
However, it is important to emphasize \ldots\ that the primitive and
fundamental notions of {\it individual\/} preference and belief will
typically provide the starting point for {\it interpersonal\/}
communication and reporting processes. \ldots\ [W]e shall therefore
often be concerned to identify and examine features of the individual
learning process which relate to interpersonal issues, such as the
conditions under which an approximate consensus of beliefs might
occur in a population of individuals.
\eq
Following that theme is the key to understanding quantum tomography
from a Bayesian point of view.

The offending classical concept is an ``unknown probability,'' an
oxymoron for the precisely same reason as an unknown quantum
state.\,\footnote{If a quantum state is nothing more than a
compendium of probabilities---Bayesian probabilities---then, of
course, it is for precisely the same reason.} The procedure analogous
to quantum-state tomography is the estimation of an unknown
probability from the results of repeated trials on ``identically
prepared systems,'' all of which are said to be described by the
same, but unknown probability.

Let us first consider a trivial example to clinch the idea of
subjective probabilities before moving to a full tomographic setting.
Suppose a die is thrown 10 times. Let $k\in\{1,\ldots,6\}$ represent
the outcome of a single throw of the die, and suppose the
results in the 10 throws were these:\\

{}$k=1\;\;$  appeared 1 times,

{}$k=2\;\;$ appeared 4 times,

{}$k=3\;\;$ appeared 2 times,

{}$k=4\;\;$ appeared 2 times,

{}$k=5\;\;$ appeared 1 times,

{}$k=6\;\;$ appeared 0 times.\\

\noindent A typical inference problem is to assign a probability, $p$,
to the outcome $k=6$ in the next throw of the die, given these data.
But clearly the answer cannot be determined by the data alone. It
depends on a prior probability assignment. Possibilities include:
\begin{enumerate}
\item The assumption that the die is fair. In this case $p$ is
independent of the data and given by $p=1/6$.

\item A totally uninformative prior. In this case, a generalization
of the Laplace rule of succession \cite{Jaynes2003} gives $p=1/16$.

\item The die came from a box that contains only trick dice of two
types: The first type never comes up 1, and the other type never
comes up 6. In this case, $p=0$.
\end{enumerate}
The point of this trivial example cannot be stressed enough.  Data
alone is never enough to specify a probability distribution.  The
only way to ``derive'' a probability distribution from data is to
cheat---to make use of an implicit probability assignment made prior
to the collection of the data. This holds as much for a problem like
the present one, whose setting is a finite number of trials, as for a
problem containing a potentially infinite number of trials.

The way to eliminate unknown probabilities from the discussion of
infinite numbers of trials was introduced by Bruno de Finetti in the
early 1930s \cite{DeFinetti1990,DeFinettiCollected}.  The idea is to
make explicit the implicit class of priors generally used in such
problems.  He did this by focusing on the meaning of the {\it
equivalence\/} of repeated trials. What could equivalent trials
mean---de Finetti asked---but that a probability assignment for
multiple trials should be symmetric under the permutation of those
trials? With his {\it classical representation theorem}, de Finetti
\cite{DeFinetti1990} showed that a multi-trial probability assignment
that is permutation-symmetric for an arbitrarily large number of
trials---he called such multi-trial probabilities {\it
exchangeable\/}---is equivalent to a probability for the ``unknown
probabilities.''  Thus the unsatisfactory concept of an unknown
probability vanishes from the description in favor of the fundamental
idea of assigning an exchangeable probability distribution to
multiple trials.

This cue in hand, it is easy to see how to reword the description
of quantum-state tomography to meet our goals.  What is relevant
is simply a judgment on the part of the experimenter---notice the
essential subjective character of this ``judgment''---that there
is no distinction between the systems the device is preparing.  In
operational terms, this is the judgment that {\it all the systems
are and will be the same as far as observational predictions are
concerned}.  At first glance this statement might seem to be
contentless, but the important point is this: To make this
statement, one need never use the notion of an unknown state---a
completely operational description is good enough. Putting it into
technical terms, the statement is that if the experimenter judges
a collection of $N$ of the device's outputs to have an overall
quantum state $\rho^{(N)}$, he will also judge any permutation of
those outputs to have the same quantum state $\rho^{(N)}$.
Moreover, he will do this no matter how large the number $N$ is.
This, complemented only by the consistency condition that for any
$N$ the state $\rho^{(N)}$ be derivable from $\rho^{(N+1)}$, makes
for the complete story.

The words ``quantum state'' appear in this formulation, just as in
the original formulation of tomography, but there is no longer any
mention  of {\it unknown\/} quantum states.  The state $\rho^{(N)}$
is known by the experimenter (if no one else), for it represents his
state of belief.  More importantly, the experimenter is in a position
to make an unambiguous statement about the structure of the whole
sequence of states $\rho^{(N)}$: Each of the states $\rho^{(N)}$ has
a kind of permutation invariance over its factors. The content of the
{\it quantum de Finetti representation
theorem}~\cite{Hudson1976,Hudson1981}---which we will demonstrate in
a later section---is that a sequence of states $\rho^{(N)}$ can have
these properties, which are said to make it an {\it exchangeable\/}
sequence, if and only if each term in it can also be written in the
form
\begin{equation}
\rho^{(N)}=\int P(\rho)\, \rho^{\otimes N}\, d\rho\;,
\label{Jeremy}
\end{equation}
where $\rho^{\otimes N}= \rho\otimes\rho\otimes\cdots\otimes\rho$ is
an $N$-fold tensor product and $P(\rho)$ is a fixed probability
distribution over density operators.\,\footnote{For further technical
elaborations on the quantum de Finetti theorem, see
Ref.~\cite{DeFinettiLump}.}

The interpretive import of this theorem is paramount. It alone gives
a mandate to the term unknown state in the usual description of
tomography.  It says that the experimenter can act {\it as if\/} his
state of belief $\rho^{(N)}$ comes about because he knows there is a
``man in the box,'' hidden from view, repeatedly preparing the same
state $\rho$.  He does not know which such state, and the best he can
say about the unknown state is captured in the probability
distribution $P(\rho)$.

The quantum de Finetti theorem furthermore makes a connection to the
overarching theme of Bayesianism stressed above.  It guarantees for
two independent observers---as long as they have a rather minimal
agreement in their initial beliefs---that the outcomes of a
sufficiently informative set of measurements will force a
convergence in their state assignments for the remaining
systems~\cite{Schack2000}.  This ``minimal'' agreement is
characterized by a judgment on the part of both parties that the
sequence of systems is exchangeable, as described above, and a
promise that the observers are not absolutely inflexible in their
opinions.  Quantitatively, the latter means that though $P(\rho)$
might be arbitrarily close to zero, it can never vanish.

This coming to agreement works because an exchangeable density
operator sequence can be updated to reflect information gathered from
measurements by a quantum version of Bayes's rule for updating
probabilities. Specifically, suppose the starting point of a quantum
tomography experiment is a prior state
\begin{equation}
\rho^{(N+M)} = \int { P(\rho)}\, \rho^{\otimes(N+M)} d\rho
\end{equation}
for $N+M$ copies of the system. Then the first $N$ systems are
measured. Say the measurement outcomes are represented by a vector
$\bbox{\alpha}=(\alpha_1,\ldots,\alpha_N)$. It can be shown that the
post-measurement state of the remaining $M$ copies conditioned on the
outcome $\bbox{\alpha}$ is of the form
\begin{equation}
\rho^{(M)} = \int P(\rho|\bbox{\alpha})\, \rho^{\otimes M} d\rho\;,
\end{equation}
where $P(\rho|\bbox{\alpha})$ is given by a quantum Bayes
rule~\cite{Schack2000}. In the special case that the same
measurement, $\{E_\alpha\}$, is measured on all $N$ copies, the
quantum Bayes rule takes the simple form
\begin{equation}
P(\rho|\bbox{\alpha})={P(\rho){ P(\bbox{\alpha}|\rho)} \over
P(\alpha)} \;,
\end{equation}
where
\begin{equation}
P(\bbox{\alpha}|\rho) = \tr\big(\,\rho^{\otimes N}\,
    E_{\alpha_1}\otimes\cdots\otimes E_{\alpha_N}\big)
\end{equation}
and
\begin{equation}
P(\bbox{\alpha}) = \int P(\rho) P(\bbox{\alpha}|\rho)\, d\rho \;.
\end{equation}
For a sufficiently informative set of measurements, as $N$ becomes
large, the updated probability $P(\rho|\bbox{\alpha})$ becomes highly
peaked on a particular state $\rho_{\alpha}$ dictated by the
measurement results, regardless of the prior probability $P(\rho)$,
as long as $P(\rho)$ is nonzero in a neighborhood of $\rho_{\alpha}$.
Suppose the two observers have different initial beliefs,
encapsulated in different priors $P_i(\rho)$, $i=1,2$.  The
measurement results force them with high probability to a common
state of belief in which any number $M$ of additional systems are
assigned the product state $\rho_{\alpha}^{\otimes M}$, i.e.,
\begin{equation}
\int P_i(\rho|\bbox{\alpha})\,\rho^{\otimes M}\,d\rho
\quad{\longrightarrow}\quad
\rho_{\alpha}^{\otimes M}
\label{HannibalLecter}
\end{equation}
for $N$ sufficiently large.

This shifts the perspective on the purpose of quantum-state
tomography:  It is not about uncovering some ``unknown state of
nature,'' but rather about the various observers' coming to agreement
over future probabilistic predictions.\,\footnote{For an emphasis of
this point in the setting of quantum cryptography, see
Ref.~\cite{Fuchs2000b}.} In this connection, it is interesting to
note that the quantum de Finetti theorem and the conclusions just
drawn from it work only within the framework of complex vector-space
quantum mechanics. For quantum mechanics based on real and
quaternionic Hilbert spaces, the connection between exchangeable
density operators and unknown quantum states does not
hold~\cite{Caves2002b}.

The plan of the remainder of the paper is as follows. In
Sec.~\ref{sec-classical}, we discuss the classical de Finetti
representation theorem~\cite{DeFinetti1990,Heath1976} in the context
of Bayesian probability theory.  In Sec.~\ref{sec-quantum}, we
introduce the Bayesian formulation of tomography in terms of
exchangeable multi-system density operators, accompanied by a
critical discussion of objectivist formulations of tomography.
Furthermore, we state the quantum-state de Finetti representation
theorem. Section~\ref{sec-proof} presents an elementary proof of the
quantum de Finetti theorem. There, also, we introduce a novel
measurement technique for tomography based upon generalized quantum
measurements. In Sec.~\ref{sec-intermezzo}, we come to an intermezzo,
mentioning possible extensions of the main theorem. In
Sec.~\ref{Emma}, we change course to consider the issue of quantum
operations.  In particular, we argue that (at least some) quantum
operations should be considered subjective states of belief, just as
quantum states themselves. This brings to the fore the issue of
``unknown quantum operations'' within a Bayesian formulation of
quantum mechanics. In Sec.~\ref{Katie}, we pose the need for a
version of a quantum de Finetti theorem for quantum operations in
order to make sense of quantum-process tomography from a Bayesian
point of view. In Sec.~\ref{sectheorem}, we make the statement of the
theorem precise. And in Sec.~\ref{secproof}, we run through the proof
of this quantum-process de Finetti theorem. Finally in
Sec.~\ref{Kiki}, we conclude with a discussion of where this research
program is going. In particular, we defend ourselves against the
(glib) ``shot gun'' reaction that all of this amounts to a rejection
of realism altogether:  It simply does not.

\section{The Classical de Finetti Theorem} \label{sec-classical}

The tension between the objectivist and Bayesian points of view is
not new with quantum mechanics.  It arises already in classical
probability theory in the form of the war between ``objective'' and
``subjective'' interpretations~\cite{Daston1994}. According to the
subjective or Bayesian interpretation, probabilities are measures of
personal belief, reflecting how an agent would behave or bet in a
certain situation. On the other hand, the objective
interpretations---in all their varied forms, from frequency
interpretations to propensity interpretations---attempt to view
probabilities as real states of affairs or ``states of nature'' that
have nothing to do with an agent at all.  Following our discussion in
Sec.~\ref{sec-intro}, it will come as no surprise to the reader that
the authors wholeheartedly adopt the Bayesian approach. For us, the
reason is simply our experience with this question, part of which is
an appreciation that objective interpretations inevitably run into
insurmountable difficulties. (See
Refs.~\cite{DeFinetti1989,Savage1972,Bernardo1994,Kyburg1980} for a
sampling of criticisms of the objectivist approach.)

We will note briefly, however, that the game of roulette provides an
illuminating example. In the European version of the game, the
possible outcomes are the numbers $0,1,\ldots,36$.  For a player
without any privileged information, all 37 outcomes have the same
probability $p=1/37$.  But suppose that shortly after the ball is
launched by the croupier, another player obtains information about
the ball's position and velocity relative to the wheel. Using the
information obtained, this other player can make more accurate
predictions than the first.\,\footnote{An entertaining account of a
serious attempt to make money from this idea can be found in
Ref.~\cite{NewtonianCasino}.} His probability is peaked around some
group of numbers. The probabilities are thus different for two
players with different states of belief.

Whose probability is the true probability? {}From the Bayesian
viewpoint, this question is meaningless:  There is no such thing as a
true probability.  All probability assignments are subjective
assignments based specifically upon one's prior data and beliefs.

For sufficiently precise data---including precise initial data on
positions and velocities and probably also including other details
such as surface properties of the wheel---Newtonian mechanics assures
us that the outcome can be predicted with certainty.  This is an
important point: The determinism of classical physics provides a
strong reason for adopting the subjectivist view of
probabilities~\cite{Giere1973}.  If the conditions of a trial are
exactly specified, the outcomes are predictable with certainty, and
all probabilities are 0 or 1. In a deterministic theory, all
probabilities strictly greater than 0 and less than 1 arise as a
consequence of incomplete information and depend upon their
assigner's state of belief.

Of course, we should keep in mind that our ultimate goal is to
consider the status of quantum states and, by way of them, quantum
probabilities. One can ask, ``Does this not change the flavor of
these considerations?''
Quantum mechanics is avowedly {\it not\/} a theory of one's ignorance of
a set of hidden variables~\cite{BellBook}:\,\footnote{Perhaps, more
carefully, we should have added, ``without a stretch of the
imagination.''  For a stretch of the imagination, see
Ref.~\cite{GoldsteinBook}.} So how can its probabilities be subjective? In
Sec.~\ref{sec-quantum} we argue that despite the intrinsic
indeterminism of quantum mechanics, the essence of the above discussion
carries over to the quantum setting intact. Furthermore, there are
specifically quantum-motivated arguments for a Bayesian
interpretation of quantum probabilities.

For the present, though, let us consider in some detail the general
problem of a repeated experiment---spinning a roulette wheel $N$
times is an example. As discussed briefly in Sec.~\ref{sec-intro},
this allows us to make a conceptual connection to quantum-state
tomography.  Here the individual trials are described by discrete
random variables $x_n\in\{1,2,\ldots,k\}$, $n=1,\ldots,N$; that is to
say, there are $N$ random variables, each of which can assume $k$
discrete values. In an objectivist theory, such an experiment has a
standard formulation in which the probability in the multi-trial
hypothesis space is given by an independent, identically distributed
distribution
\begin{equation}
p(x_1,x_2,\ldots,x_N)\,=\,p_{x_1} p_{x_2} \cdots p_{x_N}\, =\,
p_1^{n_{\scriptscriptstyle 1}} p_2^{n_{\scriptscriptstyle 2}}\cdots
p_k^{n_{\scriptscriptstyle k}}\;,
\label{eq-iid}
\end{equation}
where $n_j$ is the number of times outcome $j$ is listed in the
vector $(x_1,x_2,\ldots,x_N)$, so that $\sum_j n_j=N$. The number
$p_j$ ($j=1,\ldots,k$) describes the objective, ``true'' probability
that the result of a single experiment will be $j$ ($j=1,\ldots,k$).
This simple description---for the objectivist---only describes the
situation from a kind of ``God's eye'' point of view.  To the
experimentalist, the ``true'' probabilities $p_1,\ldots,p_k$ will
very often be {\it unknown\/} at the outset.  Thus, his burden is to
estimate the unknown probabilities by a statistical analysis of the
experiment's outcomes.

In the Bayesian approach, it does not make sense to talk about
estimating a true probability.  Instead, a Bayesian assigns a prior
probability distribution $p(x_1,x_2,\ldots,x_N)$ on the multi-trial
hypothesis space and then uses Bayes's theorem to update the
distribution in the light of measurement results. A common criticism
from the objectivist camp is that the choice of distribution
$p(x_1,x_2,\ldots,x_N)$ with which to start the process seems overly
arbitrary to them. On what can it be grounded, they would ask? {}From
the Bayesian viewpoint, the subjectivity of the prior is a strength
rather than a weakness, because assigning a prior amounts to laying
bare the necessarily subjective assumptions behind {\it any\/}
probabilistic argument, be it Bayesian or objectivist. Choosing a
prior among all possible distributions on the multi-trial hypothesis
space is, however, a daunting task. As we will now see, this task
becomes tractable by the de Finetti representation theorem.

It is very often the case that one or more features of a problem
stand out so clearly that there is no question about how to
incorporate them into an initial assignment. In the present case,
the key feature is contained in the assumption that an arbitrary
number of repeated trials are equivalent.  This means that one has
no reason to believe there will be a difference between one trial
and the next. In this case, the prior distribution is judged to
have the sort of permutation symmetry discussed briefly in
Sec.~\ref{sec-intro}, which de Finetti \cite{DeFinettiCollected}
called {\it exchangeability}.  The rigorous definition of
exchangeability proceeds in two stages.

A probability distribution $p(x_1,x_2,\ldots,x_N)$ is said to be
{\it symmetric\/} (or finitely exchangeable) if it is invariant
under permutations of its arguments, i.e., if
\begin{equation}
p\bigl(x_{\pi(1)},x_{\pi(2)},\ldots,x_{\pi(N)}\bigr) =
p(x_1,x_2,\ldots,x_N)
\end{equation}
for any permutation $\pi$ of the set $\{1,\ldots,N\}$. The
distribution $p(x_1,x_2,\ldots,x_N)$ is called {\it
exchangeable\/} (or infinitely exchangeable) if it is symmetric
and if for any integer $M>0$, there is a symmetric distribution
$p_{N+M}(x_1,x_2,\ldots,x_{N+M})$ such that
\begin{equation}
p(x_1,x_2,\ldots,x_N)\; =
\sum_{x_{N+1},\ldots,x_{N+M}}
p_{N+M}(x_1,\ldots,x_N,x_{N+1},\ldots,x_{N+M})
\;.
\label{eq-marginal}
\end{equation}
This last statement means that the distribution $p$ can be extended
to a symmetric distribution of arbitrarily many random variables.
Expressed informally, an exchangeable distribution can be thought of
as arising from an infinite sequence of random variables whose order
is irrelevant.

We now come to the main statement of this section: If a probability
distribution $p(x_1,x_2,\ldots,x_N)$ is exchangeable, then it can be
written uniquely in the form
\begin{equation}
p(x_1,x_2,\ldots,x_N)=\int_{{\cal S}_k} P(\vec{p})\,p_{x_1}
p_{x_2} \cdots p_{x_N}\,d\vec{p}
=
\int_{{\cal S}_k} P(\vec{p})\, p_1^{n_{\scriptscriptstyle 1}}
p_2^{n_{\scriptscriptstyle 2}}\cdots p_k^{n_{\scriptscriptstyle k}} \,
d\vec{p}\;,
\label{eq-repr}
\end{equation}
where $\vec{p}=(p_1,p_2,\ldots,p_k)$, and the integral is taken over
the probability simplex
\begin{equation}
{\cal S}_k=\left\{\vec{p}\mbox{ : }\; p_j\ge0\mbox{ for all } j\mbox{
and } \sum_{j=1}^k p_j=1\right\}.
\end{equation}
Furthermore, the function $P(\vec{p})\ge0$ is required to be a
probability density function on the simplex:
\begin{equation}
\int_{{\cal S}_k} P(\vec{p})\,d\vec{p}=1\;.
\end{equation}
Equation~(\ref{eq-repr}) comprises the classical de Finetti
representation theorem for discrete random variables. (A simple proof
of this theorem in the case of binary random variables can be found
in Refs.~\cite{Heath1976,Caves2002b}.)

Let us reiterate the importance of this result for the present
considerations.  It says that an agent, making solely the judgment of
exchangeability for a sequence of random variables $x_j$, can proceed
{\it as if\/} his state of belief had instead come about through
ignorance of an {\it unknown}, but objectively existent set of
probabilities $\vec{p}$.  His precise ignorance of $\vec{p}$ is
captured by the ``probability on probabilities'' $P(\vec{p})$.  This
is in direct analogy to what we desire of a solution to the problem
of the unknown quantum state in quantum-state tomography.

As a final note before finally addressing the quantum problem in
Sec.~\ref{sec-quantum}, we point out that both conditions in the
definition of exchangeability are crucial for the proof of the de
Finetti theorem.  In particular, there are probability
distributions $p(x_1,x_2,\ldots,x_N)$ that are symmetric, but not
exchangeable.  A simple example is the distribution $p(x_1,x_2)$
of two binary random variables $x_1,x_2\in\{0,1\}$,
\begin{eqnarray}
&& p(0,0) = p(1,1) = 0\;,
\label{HocusPocus}
\\
&& p(0,1) = p(1,0) = \frac{1}{2} \;.
\label{Hiroshima}
\end{eqnarray}
One can easily check that $p(x_1,x_2)$ cannot be written as the
marginal of a symmetric distribution of three variables, as in
Eq.~(\ref{eq-marginal}). Therefore it can have no representation
along the lines of Eq.~(\ref{eq-repr}).  (For an extended discussion
of this, see Ref.~\cite{Jaynes1986}.)  Indeed,
Eqs.~(\ref{HocusPocus}) and (\ref{Hiroshima}) characterize a perfect
``anticorrelation'' of the two variables, in contrast to the positive
correlation implied by distributions of de Finetti form.

\section{The quantum de Finetti representation} \label{sec-quantum}

Let us now return to the problem of quantum-state tomography
described in Sec.~\ref{sec-intro}. In the objectivist formulation of
the problem, a device repeatedly prepares copies of a system in the
same quantum state $\rho$. This is generally a mixed-state density
operator on a Hilbert space ${\cal H}_d$ of $d$ dimensions. We call
the totality of such density operators ${\cal D}_d$.  The joint
quantum state of the $N$ systems prepared by the device is then
given by
\begin{equation}
\rho^{\otimes N}=\rho\otimes\rho\otimes\cdots\otimes\rho \;,
\end{equation}
the $N$-fold tensor product of $\rho$ with itself. This, of
course, is a very restricted example of a density operator on the
tensor-product Hilbert space ${\cal H}_d^{\otimes N}\equiv {\cal
H}_d\otimes\cdots\otimes{\cal H}_d$. The experimenter, who
performs quantum-state tomography, tries to determine $\rho$ as
precisely as possible. Depending upon the version of the argument,
$\rho$ is interpreted as the ``true'' state of each of the systems
or as a description of the ``true'' preparation procedure.

We have already articulated our dissatisfaction with this way of
stating the problem, but we give here a further sense of why both
interpretations above are untenable.  Let us deal first with the
version where $\rho$ is regarded as the true, objective state of
each of the systems.  In this discussion it is useful to consider
separately the cases of mixed and pure states $\rho$.  The
arguments against regarding mixed states as objective properties
of a quantum system are essentially the same as those against
regarding probabilities as objective. In analogy to the roulette
example given in the previous section, we can say that, whenever
an observer assigns a mixed state to a physical system, one can
think of another observer who assigns a different state based on
privileged information.

The quantum argument becomes yet more compelling if the apparently
nonlocal nature of quantum states is taken into consideration.
Consider two parties, $A$ and $B$, who are far apart in space, say
several light years apart. Each party possesses a spin-1/2 particle.
Initially the joint state of the two particles is the maximally
entangled pure state
${1\over\sqrt2}(|0\rangle|0\rangle+|1\rangle|1\rangle)$.
Consequently, $A$ assigns the totally mixed state
${1\over2}(|0\rangle\langle0|+|1\rangle\langle1|)$ to her own
particle. Now $B$ makes a measurement on his particle, finds the
result 0, and assigns to $A$'s particle the pure state $|0\rangle$.
Is this now the ``true,'' objective state of $A$'s particle? At what
precise time does the objective state of $A$'s particle change from
totally mixed to pure?  If the answer is ``simultaneously with $B$'s
measurement,'' then what frame of reference should be used to
determine simultaneity?  These questions and potential paradoxes are
avoided if states are interpreted as states of belief. In our
example, $A$ and $B$ have different states of belief and therefore
assign different states. For a detailed analysis of this example, see
Ref.~\cite{Peres-9906a}; for an experimental investigation see
Ref.~\cite{Scarani2000}.

If one admits that mixed states cannot be objective properties,
because another observer, possessing privileged information, can know
which pure state underlies the mixed state, then it becomes very
tempting to regard the pure states as giving the ``true'' state of a
system.  Probabilities that come from pure states would then be
regarded as objective, and the probabilities for pure states within
an ensemble decomposition of mixed states would be regarded as
subjective, expressing our ignorance of which pure state is the
``true'' state of the system.  An immediate and, in our view,
irremediable problem with this idea is that a mixed state has
infinitely many ensemble decompositions into pure states
\cite{Schrodinger1936,Jaynes1957b,Hughston1993}, so the distinction
between subjective and objective becomes hopelessly blurred.

This problem can be made concrete by the example of a
spin-${1\over2}$ particle. Any pure state of the particle can be
written in terms of the Pauli matrices as
\begin{equation}  \label{eq-poincare}
|\vec n\rangle\langle\vec n|={1\over2}(I+{\vec n}\cdot\bbox{\sigma})
={1\over2}(I+n_1\sigma_1+n_2\sigma_2+n_3\sigma_3)\;,
\end{equation}
where the unit vector ${\vec n}=n_1\vec e_1+n_2\vec e_2+n_3\vec
e_3$ labels the pure state, and $I$ denotes the unit operator.  An
arbitrary state $\rho$, mixed or pure, of the particle can be
expressed as
\begin{equation}
\rho={1\over2}(I+\vec S\cdot\bbox{\sigma}) \;,
\label{eq-rhoqubit}
\end{equation}
where $0\le|\vec S|\le1$.  If $|\vec S|<1$, there is an infinite
number of ways in which $\vec S$ can be written in the form $\vec
S=\sum_j p_j{\vec n}_j$, $|\vec n_j|=1$, with the numbers $p_j$
comprising a probability distribution, and hence an infinite number
of ensemble decompositions of $\rho$:
\begin{equation}
\rho = \sum_jp_j{1\over2}(I+{\vec n}_j\cdot\bbox{\sigma})
=\sum_j p_j|\vec n_j\rangle\langle\vec n_j|\;.
\label{eq-decomp}
\end{equation}

Suppose for specificity that the particle's state is a mixed state
with $\vec S={1\over2}\,\vec e_3$.  Writing $\vec S={3\over4}\vec
e_3+{1\over4}(-\vec e_3)$ gives the eigendecomposition,
\begin{equation}
\rho=
{3\over4}|\vec e_3\rangle\langle\vec e_3|
+{1\over4}|\mathord{-}\vec e_3\rangle\langle\mathord{-}\vec e_3|\;,
\end{equation}
where we are to regard the probabilities $3/4$ and $1/4$ as
subjective expressions of ignorance about which eigenstate is the
``true'' state of the particle.  Writing $\vec S={1\over2}\vec
n_++{1\over2}\vec n_-$, where $\vec n_{\pm}={1\over2}\vec
e_3\pm{\sqrt3\over2}\vec e_1$, gives another ensemble decomposition,
\begin{equation}
\rho=
{1\over2}|\vec n_+\rangle\langle\vec n_+|
+{1\over2}|\vec n_-\rangle\langle\vec n_-|\;,
\label{Eleanor}
\end{equation}
where we are now to regard the two probabilities of $1/2$ as
expressing ignorance of whether the ``true'' state is $|\vec
n_+\rangle$ or $|\vec n_-\rangle$.

The problem becomes acute when we ask for the probability that a
measurement of the $z$ component of spin yields spin up; this
probability is given by $\langle\vec e_3|\rho|\vec
e_3\rangle={1\over2}(1+{1\over2}\langle\vec e_3|\sigma_3|\vec
e_3\rangle)=3/4$.  The eigendecomposition gets this probability by
the route
\begin{equation}
\langle\vec e_3|\rho|\vec e_3\rangle=
{3\over4}
\underbrace{|\langle\vec e_3|\vec e_3\rangle|^2}_%
{\displaystyle{1}}
+{1\over4}
\underbrace{|\langle\vec e_3|\mathord{-}\vec e_3\rangle|^2}_%
{\displaystyle{0}}\;.
\end{equation}
Here the objective quantum probabilities, calculated from the
eigenstates, report that the particle definitely has spin up or
definitely has spin down; the overall probability of $3/4$ comes
from mixing these objective probabilities with the subjective
probabilities for the eigenstates.  The
decomposition~(\ref{Eleanor}) gets the same overall probability by
a different route,
\begin{equation}
\langle\vec e_3|\rho|\vec e_3\rangle=
{1\over2}
\underbrace{|\langle\vec e_3|\vec n_+\rangle|^2}_%
{\displaystyle{3/4}}
+{1\over2}
\underbrace{|\langle\vec e_3|\vec n_-\rangle|^2}_%
{\displaystyle{3/4}}
\;.
\end{equation}
Now the quantum probabilities tell us that the objective probability
for the particle to have spin up is $3/4$.  This simple example
illustrates the folly of trying to have two kinds of probabilities in
quantum mechanics.  The lesson is that if a density operator is even
partially a reflection of one's state of belief, the multiplicity of
ensemble decomposition means that a pure state must also be a state
of belief.

Return now to the second version of the objectivist formulation of
tomography, in which the experimenter is said to be using
quantum-state tomography to determine an unknown preparation
procedure. Imagine that the tomographic reconstruction results in the
mixed state $\rho$, rather than a pure state, as in fact all actual
laboratory procedures do.   Now there is a serious problem, because a
mixed state does not correspond to a well-defined procedure, but is
itself a probabilistic mixture of well-defined procedures, i.e., pure
states.  The experimenter is thus trying to determine an unknown
procedure that has no unique decomposition into well defined
procedures.  Thus he cannot be said to be determining an unknown
procedure at all.  This problem does not arise in a Bayesian
interpretation, according to which all quantum states, pure or mixed,
are states of belief.  In analogy to the classical case, the quantum
de Finetti representation provides an operational definition for the
idea of an unknown quantum state in this case.

Let us therefore turn to the Bayesian formulation of the
quantum-state tomography problem. Before the tomographic
measurements, the Bayesian experimenter assigns a prior quantum state
to the joint system composed of the $N$ systems, reflecting his prior
state of belief.  Just as in the classical case, this is a daunting
task unless the assumption of exchangeability is justified.

The definition of the quantum version of exchangeability is
closely analogous to the classical definition.  Again, the
definition proceeds in two stages.  First, a joint state
$\rho^{(N)}$ of $N$ systems is said to be {\it symmetric\/} (or
finitely exchangeable) if it is invariant under any permutation of
the systems.  To see what this means formally, first write out
$\rho^{(N)}$ with respect to any orthonormal tensor-product basis
on ${\cal H}_d^{\otimes N}$, say
$|i_1\rangle|i_2\rangle\cdots|i_N\rangle$, where
$i_k\in\{1,2,\ldots,d\}$ for all $k\,$.  The joint state takes the
form
\begin{equation}
\rho^{(N)}=\sum_{i_1,\ldots,i_N;j_1,\ldots,j_N}
R^{(N)}_{i_1,\ldots,i_N;j_1,\ldots,j_N}\,
|i_1\rangle\cdots|i_N\rangle \langle j_1| \cdots\langle j_N|\;,
\end{equation}
where $R^{(N)}_{i_1,\ldots,i_N;j_1,\ldots,j_N}$ is the density
matrix in this representation.  What we demand is that for any
permutation $\pi$ of the set $\{1,\ldots,N\}$,
\be
\rho^{(N)}=\sum_{i_1,\ldots,i_N;j_1,\ldots,j_N}
R^{(N)}_{i_{\pi(1)},\ldots,i_{\pi(N)};j_{\pi(1)},\ldots,j_{\pi(N)}}\,
|i_1\rangle\cdots|i_N\rangle \langle j_1| \cdots\langle j_N| \;,
\ee
which is equivalent to
\begin{equation}
R^{(N)}_{i_{\pi(1)},\ldots,i_{\pi(N)};j_{\pi(1)},\ldots,j_{\pi(N)}}
=R^{(N)}_{i_1,\ldots,i_N;j_1,\ldots,j_N}\;.
\end{equation}

The state $\rho^{(N)}$ is said to be {\it exchangeable\/} (or
infinitely exchangeable) if it is symmetric and if, for any $M>0$,
there is a symmetric state $\rho^{(N+M)}$ of $N+M$ systems such that
the marginal density operator for $N$ systems is $\rho^{(N)}$, i.e.,
\begin{equation}
\rho^{(N)} = \tr_M\,\rho^{(N+M)} \;,
\label{HoundDog}
\end{equation}
where the trace is taken over the additional $M$ systems.  In
explicit basis-dependent notation, this requirement is
\begin{equation}
\rho^{(N)}=
\!\!\sum_{i_1,\ldots,i_N;j_1,\ldots,j_N}
\!\!\left(\,\sum_{i_{N+1},\ldots,i_{N+M}}\!\!
R^{(N+M)}_{i_1,\ldots,i_N,i_{N+1},\ldots,i_{N+M};
j_1,\ldots,j_N,i_{N+1},\ldots,i_{N+M}}\right)\!
|i_1\rangle\cdots|i_N\rangle \langle j_1| \cdots\langle j_N|\;.
\end{equation}
In analogy to the classical case, an exchangeable density operator can
be thought of informally as the description of a subsystem of an
infinite sequence of systems whose order is irrelevant.

The precise statement of the quantum de Finetti representation
theorem~\cite{Hudson1976,Stormer1969} is that any exchangeable state
of $N$ systems can be written uniquely in the form
\begin{equation}
\rho^{(N)}=\int_{{\cal D}_d} P(\rho)\, \rho^{\otimes N}\, d\rho\;.
\label{eq-qdefinetti}
\end{equation}
Here $P(\rho)\ge0$ is normalized by
\begin{equation}
\int_{{\cal D}_d} P(\rho)\,d\rho=1\;,
\end{equation}
with $d\rho$ being a suitable measure on density operator space
${\cal D}_d$ [e.g., one could choose $d\rho=dS\,d\Omega$ in the
parameterization~(\ref{eq-rhoqubit}) for a spin-1/2 particle].  The
upshot of the theorem, as already advertised, is that it makes it
possible to think of an exchangeable quantum-state assignment {\it
as if\/} it were a probabilistic mixture characterized by a probability
density $P(\rho)$ for the product states $\rho^{\otimes N}$.

Just as in the classical case, both components of the definition
of exchangeability are crucial for arriving at the representation
theorem of Eq.~(\ref{eq-qdefinetti}).  The reason now, however, is
much more interesting than it was previously.  In the classical
case, extendibility was used solely to exclude anticorrelated
probability distributions. Here extendibility is necessary to
exclude the possibility of Bell inequality violations for
measurements on the separate systems. This is because the
assumption of symmetry alone for an $N$-party quantum system does
not exclude the possibility of quantum entanglement, and all
states that can be written as a mixture of product states---of
which Eq.~(\ref{eq-qdefinetti}) is an example---have no
entanglement~\cite{Bennett1996}. A simple example for a state that is
symmetric but not exchangeable is
the Greenberger-Horne-Zeilinger state of three spin-$1\over2$
particles~\cite{Mermin1990},
\begin{equation}
|\mbox{GHZ}\rangle=\frac{1}{\sqrt{2}}\Big(|0\rangle|0\rangle|0\rangle+
|1\rangle|1\rangle|1\rangle\Big)\;,
\end{equation}
which is not extendible to a symmetric state on
four systems.  This follows because the only states of four
particles that marginalize to a three-particle pure state, like
the GHZ state, are product states of the form
$|\mbox{GHZ}\rangle\langle\mbox{GHZ}|\otimes\rho$, where $\rho$ is
the state of the fourth particle; such states clearly cannot be
symmetric.  These considerations show that in order for the
proposed theorem to be valid, it must be the case that as $M$
increases in Eq.~(\ref{HoundDog}), the possibilities for
entanglement in the separate systems compensatingly
decrease~\cite{Koashi2000}.

\section{Proof of the quantum de Finetti theorem} \label{sec-proof}

To prove the quantum version of the de Finetti theorem, we rely on
the classical theorem as much as possible.  We start from an
exchangeable density operator $\rho^{(N)}$ defined on $N$ copies of a
system.  We bring the classical theorem to our aid by imagining a
sequence of identical quantum measurements on the separate systems
and considering the outcome probabilities they would produce. Because
$\rho^{(N)}$ is assumed exchangeable, such identical measurements
give rise to an exchangeable probability distribution for the
outcomes.  The trick is to recover enough information from the
statistics of these measurements to characterize the exchangeable
density operator.

With this in mind, the proof is expedited by making use of the theory
of generalized quantum measurements or positive operator-valued
measures (POVMs)~\cite{Davies1970,Kraus1983,Peres1993a}. POVMs
generalize the textbook notion of measurement by distilling the
essential properties that make the Born rule work.  The generalized
notion of measurement is this:  {\it Any\/} set ${\cal
E}=\{E_\alpha\}$ of positive-semidefinite operators on ${\cal H}_d$
that forms a resolution of the identity, i.e., that satisfies
\begin{equation}
\langle\psi|E_\alpha|\psi\rangle\ge0\,,\quad\mbox{for all
$|\psi\rangle\in{\cal H}_d$}
\label{Hank}
\end{equation}
and
\begin{equation}
\sum_\alpha E_\alpha = I\;,
\label{Hannibal}
\end{equation}
corresponds to at least one laboratory procedure counting as a
measurement. The outcomes of the measurement are identified with the
indices $\alpha$, and the probabilities of those outcomes are
computed according to the generalized Born rule,
\begin{equation}
p_\alpha=\tr\big(\rho E_\alpha\big) \;.
\end{equation}
The set ${\cal E}$ is called a POVM, and the operators $E_\alpha$ are
called POVM elements. Unlike standard or von Neumann measurements,
there is no limitation on the number of values $\alpha$ can take, the
operators $E_\alpha$ need not be rank-1, and there is no requirement
that the $E_\alpha$ be idempotent and mutually orthogonal.  This
definition has important content because the older notion of
measurement is simply too restrictive: there are laboratory
procedures that clearly should be called ``measurements,'' but that
cannot be expressed in terms of the von Neumann measurement process
alone.

One might wonder whether the existence of POVMs contradicts
everything taught about standard measurements in the traditional
graduate textbooks~\cite{QuantumClassics1} and the well-known
classics~\cite{QuantumClassics2}. It does not. The reason is that
any POVM can be represented formally as a standard measurement on an
ancillary system that has interacted in the past with the system of
main interest.  Thus in a certain sense, von Neumann measurements
capture everything that can be said about quantum measurements
\cite{Kraus1983}. A way to think about this is that by learning
something about the ancillary system through a standard measurement,
one in turn learns something about the system of real interest.
Indirect though this might seem, it can be a very powerful
technique, sometimes revealing information that could not have been
revealed otherwise~\cite{Holevo1973}.

For instance, by considering POVMs, one can consider measurements
with an outcome cardinality that exceeds the dimensionality of the
Hilbert space. What this means is that whereas the statistics of a
von Neumann measurement can only reveal information about the $d$
diagonal elements of a density operator $\rho$, through the
probabilities ${\rm tr}\big(\rho\Pi_i\big)$, the statistics of a
POVM generally can reveal things about the off-diagonal elements,
too. It is precisely this property that we take advantage of in our
proof of the quantum de Finetti theorem.

Our problem hinges on finding a special kind of POVM, one for which
any set of outcome probabilities specifies a unique operator.  This
boils down to a problem of pure linear algebra. The space of
operators on ${\cal H}_d$ is itself a linear vector space of
dimension $d^{\,2}$. The quantity ${\rm tr}(A^\dagger B)$ serves as
an inner product on that space. If the POVM elements $E_\alpha$ span
the space of operators---there must be at least $d^{\,2}$ POVM
elements in the set---the measurement probabilities $p_\alpha={\rm
tr}\big(\rho E_\alpha\big)$---now thought of as {\it projections\/}
in the directions $E_\alpha$---are sufficient to specify a unique
operator $\rho$.  Two distinct density operators $\rho$ and $\sigma$
must give rise to different measurement statistics. Such
measurements, which might be called {\it informationally complete},
have been studied for some time~\cite{Prugovecki1977}.

For our proof we need a slightly refined notion---that of a {\it
minimal\/} informationally complete measurement.  If an
informationally complete POVM has more than $d^{\,2}$ operators
$E_\alpha$, these operators form an overcomplete set.  This means
that given a set of outcome probabilities $p_\alpha$, there is
generally {\it no\/} operator $A$ that generates them according to
$p_\alpha={\rm tr}\big(AE_\alpha\bigr)$.  Our proof requires the
existence of such an operator, so we need a POVM that has
precisely $d^{\,2}$ linearly independent POVM elements $E_\alpha$.
Such a POVM has the minimal number of POVM elements to be
informationally complete.  Given a set of outcome probabilities
$p_\alpha$, there is a unique operator $A$ such that
$p_\alpha={\rm tr}\big(AE_\alpha\bigr)$, even though, as we
discuss below, $A$ is not guaranteed to be a density operator.

Do minimal informationally complete POVMs exist?  The answer is yes.
We give here a simple way to produce one, though there are surely
more elegant ways with greater
symmetry~\cite{D'Ariano2003,Renes2003}. Start with a complete
orthonormal basis $|e_j\rangle$ on ${\cal H}_d$, and let
$\Gamma_{jk}=|e_j\rangle\langle e_k|$.  It is easy to check that the
following $d^{\,2}$ rank-1 projectors $\Pi_\alpha$ form a linearly
independent set.
\begin{enumerate}
\item For $\alpha=1,\ldots,d$, let
\begin{equation}
\Pi_\alpha \equiv \Gamma_{jj}\,,
\end{equation}
where $j$, too, runs over the values $1,\ldots,d$.

\item For $\alpha=d+1,\ldots,\frac{1}{2}d(d+1)$, let
\begin{equation}
\Pi_\alpha \equiv \Gamma^{(1)}_{jk} =
\frac{1}{2}\Big(|e_j\rangle+|e_k\rangle\Big)
\Big(\langle e_j|+\langle e_k|\Big)
=
\frac{1}{2}(\Gamma_{jj}+\Gamma_{kk}+\Gamma_{jk}+\Gamma_{kj})\;,
\end{equation}
where $j<k$.

\item Finally, for $\alpha= \frac{1}{2}d(d+1) + 1, \ldots,d^{\,2}$, let
\begin{equation}
\Pi_\alpha \equiv \Gamma^{(2)}_{jk}
= \frac{1}{2}\Big(|e_j\rangle+i|e_k\rangle\Big)
\Big(\langle e_j|-i\langle e_k |\Big)
=\frac{1}{2}(\Gamma_{jj}+\Gamma_{kk}-i\Gamma_{jk}+i\Gamma_{kj})\;,
\end{equation}
where again $j<k$.
\end{enumerate}
All that remains is to transform these (positive-semidefinite)
linearly independent operators $\Pi_\alpha$ into a proper POVM.
This can be done by considering the positive semidefinite operator
$G$ defined by
\begin{equation}
G=\sum_{\alpha=1}^{d^2}\Pi_\alpha\;.
\label{Herbert}
\end{equation}
It is straightforward to show that $\langle\psi|G|\psi\rangle>0$ for
all $|\psi\rangle\ne0$, thus establishing that $G$ is positive
definite and hence invertible. Applying the (invertible) linear
transformation $X\rightarrow\, G^{-1/2}XG^{-1/2}$ to
Eq.~(\ref{Herbert}), we find a valid decomposition of the identity,
\begin{equation}
I=\sum_{\alpha=1}^{d^2}G^{-1/2}\Pi_\alpha G^{-1/2}\;.
\end{equation}
The operators
\begin{equation}
E_\alpha=G^{-1/2}\Pi_\alpha G^{-1/2}
\label{Knickerbocker}
\end{equation}
satisfy the conditions of a POVM, Eqs.~(\ref{Hank}) and
(\ref{Hannibal}), and moreover, they retain the rank and
linear independence of the original $\Pi_\alpha$.

It is worthwhile noting a special property of all minimal
informationally complete POVMs $\{E_\alpha\}$:  For {\it no\/}
quantum state $\rho$ is it ever the case that $\tr\rho E_\alpha=1$.
Let us show this for the case where all the $E_\alpha = k_\alpha
|\psi_\alpha\rangle\langle\psi_\alpha|$ are rank one.  Suppose it
were the case that for some $\rho$, $\tr\rho E_0=1$.  Then it would
also have to be the case that $E_0=\rho=|\psi\rangle\langle\psi|$ for
some vector $|\psi\rangle$.  But then, because $\sum E_\alpha =I$, it
follows that $\langle\psi_\alpha|\psi\rangle=0$ for all $\alpha\ne 0$.  That
is, all the $|\psi_\alpha\rangle$ must lie in a $(d-1)$-dimensional
subspace.  But then it follows that at most $(d-1)^2$ of the
operators $E_\alpha$, $\alpha\ne 0$, can be linearly independent.
Since $(d-1)^2+1\ne d^2$, we have a contradiction with the assumption
that $\{E_\alpha\}$ is an informationally complete POVM. (In
particular, for the case of the particular, minimal informationally
complete POVM in Eq.~(\ref{Knickerbocker}), it can be shown
that~\cite{Fuchs2002}
\be
P(h)\le\left[d-\frac{1}{2}\!\left(1+\cot\frac{3\pi}{4d}\right)\right]^{-1}<1\;.
\ee
For large $d$, this bound asymptotes to roughly $(0.79 d)^{-1}$.)
What this means generically is that for no informationally complete
POVM can the vectors $p_\alpha$ completely fill the probability
simplex.

\begin{figure} 
\begin{center}
\leavevmode
\epsfxsize=5cm \epsfbox{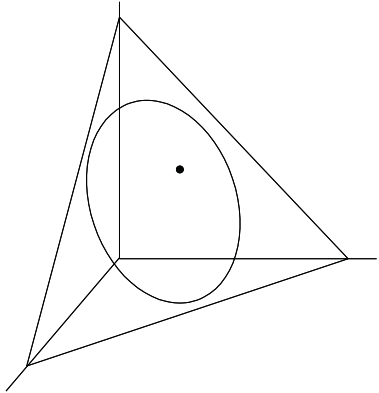} \bigskip\caption{
The planar surface represents the space of all probability
distributions over $d^2$ outcomes. If the probability distributions
are imagined to be generated by a minimal informationally complete
measurement, the set of valid quantum states in this representation
represents a convex region strictly smaller than the whole simplex.}
\end{center}
\end{figure}

With this generalized measurement (or any other one like it), we can
return to the main line of proof.  Recall we assumed that we captured
our state of belief by an exchangeable density operator $\rho^{(N)}$.
Consequently, repeated application of the (imagined) measurement
$\cal E$ must give rise to an exchangeable probability distribution
over the $N$ random variables $\alpha_n\in\{1,2,\ldots,d^{\,2}\}$,
$n=1,\ldots,N$.   We now analyze these probabilities.

Quantum mechanically, it is valid to think of the $N$ repeated
measurements of $\cal E$ as a single measurement on the Hilbert
space ${\cal H}_d^{\otimes N}\equiv {\cal
H}_d\otimes\cdots\otimes{\cal H}_d$. This measurement, which we
denote ${\cal E}^{\otimes N}$, consists of $d^{\,2N}$ POVM
elements of the form $E_{\alpha_1}\otimes\cdots\otimes
E_{\alpha_N}$. The probability of any particular outcome sequence
of length $N$, namely
$\bbox{\alpha}\equiv(\alpha_1,\ldots,\alpha_N)$, is given by the
standard quantum rule,
\begin{equation}
p^{(N)}(\bbox{\alpha})={\rm
tr}\big[\,\rho^{(N)}(E_{\alpha_1}\otimes\cdots\otimes
E_{\alpha_N})\big]\;.
\label{Humphrey}
\end{equation}
Because the distribution $p^{(N)}(\bbox{\alpha})$ is exchangeable,
we have by the classical de Finetti theorem [see
Eq.~(\ref{eq-repr})] that there exists a unique probability
density $P(\vec{p})$ on ${\cal S}_{d^2}$ such that
\begin{equation}
p^{(N)}(\bbox{\alpha})= \int_{{\cal S}_{d^2}} P(\vec{p})\,
p_{\alpha_1} p_{\alpha_2}\cdots p_{\alpha_N}\,d\vec{p}\;.
\label{Helmut}
\end{equation}

It should now begin to be apparent why we chose to imagine a
measurement $\cal E$ consisting of precisely $d^{\,2}$ linearly
independent elements. This allows us to assert the existence of a
{\it unique\/} operator $A_{\vec{p}}$ on ${\cal H}_d$
corresponding to each point $\vec{p}$ in the domain of the
integral.  The ultimate goal here is to turn Eqs.~(\ref{Humphrey})
and (\ref{Helmut}) into a single operator equation.

With that in mind, let us define $A_{\vec{p}}$ as the unique
operator satisfying the following $d^{\,2}$ linear equations:
\begin{equation}
{\rm tr}\big(A_{\vec{p}}E_\alpha\big)=
p_\alpha\;,\quad\quad\alpha=1,\ldots,d^{\,2}\;.
\label{Hamish}
\end{equation}
Inserting this definition into Eq.~(\ref{Helmut}) and manipulating it
according to the algebraic rules of tensor products---namely
$(A\otimes B)(C\otimes D)=AC\otimes BD$ and ${\rm tr}(A\otimes
B)=({\rm tr}A)({\rm tr}B)$---we see that
\begin{eqnarray}
p^{(N)}(\bbox{\alpha})
&=&
\int_{{\cal S}_{d^2}} P(\vec{p})\,{\rm
tr}\big(A_{\vec{p}}E_{\alpha_1}\big) \cdots {\rm
tr}\big(A_{\vec{p}}E_{\alpha_N}\big)\,d\vec{p}
\nonumber\\
&=&
\int_{{\cal S}_{d^2}} P(\vec{p})\,{\rm
tr}\big(A_{\vec{p}}E_{\alpha_1}\otimes \cdots\otimes
A_{\vec{p}}E_{\alpha_N}\big)\,d\vec{p}
\nonumber\\
&=&
\int_{{\cal S}_{d^2}} P(\vec{p})\,{\rm tr}\big[A_{\vec{p}}^{\otimes N}
\,
(E_{\alpha_1}\otimes\cdots\otimes E_{\alpha_N})\big]\,d\vec{p}\;.
\end{eqnarray}
If we further use the linearity of the trace, we can write the
same expression as
\begin{equation}
p^{(N)}(\bbox{\alpha})={\rm tr}\!\left[\left(\int_{{\cal S}_{d^2}}
P(\vec{p})\,A_{\vec{p}}^{\otimes n}
\,\,d\vec{p}\right)E_{\alpha_1}\otimes\cdots\otimes
E_{\alpha_N}\right].
\label{Hugo}
\end{equation}

The identity between Eqs.~(\ref{Humphrey}) and (\ref{Hugo}) must hold
for all sequences $\bbox{\alpha}$.  It follows that
\begin{equation}
\rho^{(N)}=\int_{{\cal S}_{d^2}} P(\vec{p})\,A_{\vec{p}}^{\otimes N}
\,\,d\vec{p}\;.
\label{Howard}
\end{equation}
This is because the operators $E_{\alpha_1}\otimes\cdots\otimes
E_{\alpha_N}$ form a complete basis for the vector space of
operators on ${\cal H}_d^{\otimes N}$.

Equation~(\ref{Howard}) already looks very much like our sought after
goal, but we are not there quite yet.  At this stage one has no right
to assume that the $A_{\vec{p}}$ are density operators. Indeed they
generally are not:  the integral~(\ref{Helmut}) ranges over some
points $\vec{p}$ in ${\cal S}_{d^{\,2}}$ that cannot be generated by
applying the measurement $\cal E$ to {\it any\/} quantum state. Hence
some of the $A_{\vec{p}}$ in the integral representation cannot
correspond to `unknown quantum states.'

The solution to this conundrum is provided by the overall requirement
that $\rho^{(N)}$ be a valid density operator.  This requirement
places a significantly more stringent constraint on the distribution
$P(\vec{p})$ than was the case in the classical representation
theorem.  In particular, it must be the case that $P(\vec{p})$
vanishes whenever the corresponding $A_{\vec{p}}$ is not a proper
density operator.  Let us move toward showing that.

We first need to delineate two properties of the operators
$A_{\vec{p}}$.  One is that they are Hermitian. The argument is
simply
\begin{equation}
{\rm tr}\big(E_\alpha A_{\vec{p}}^\dagger\big) = {\rm
tr}\!\left[\big(A_{\vec{p}}E_\alpha\big)^\dagger\right] = \big[{\rm
tr}\big(A_{\vec{p}}E_\alpha\big)\big]^* = {\rm
tr}\big(A_{\vec{p}}E_\alpha\big)\;,
\end{equation}
where the last step follows from Eq.~(\ref{Hamish}).  Because the
$E_\alpha$ are a complete set of linearly independent operators,
it follows that $A_{\vec{p}}^\dagger=A_{\vec{p}}$.  The second
property tells us something about the eigenvalues of
$A_{\vec{p}}$:
\begin{equation}
1=\sum_\alpha p_\alpha={\rm tr}\!\left(A_{\vec{p}}\sum_\alpha
E_\alpha\right)={\rm tr}A_{\vec{p}}\;.
\label{HepPlease}
\end{equation}
In other words the (real) eigenvalues of $A_{\vec{p}}$ must sum to
unity.

We now show that these two facts go together to imply that if there
are any non-quantum-state $A_{\vec{p}}$ with positive weight
$P(\vec{p})$ in Eq.~(\ref{Howard}), then one can find a measurement
for which $\rho^{(N)}$ produces illegal ``probabilities'' for
sufficiently large $N$.  For instance, take a particular
$A_{\vec{q}}$ in Eq.~(\ref{Howard}) that has at least one negative
eigenvalue $-\lambda<0$.  Let $|\psi\rangle$ be a normalized
eigenvector corresponding to that eigenvalue and consider the
binary-valued POVM consisting of the elements
$\widetilde{\Pi}=|\psi\rangle\langle\psi|$ and
$\Pi=I-\widetilde{\Pi}$.  Since ${\rm
tr}\big(A_{\vec{q}}\widetilde{\Pi}\big)=-\lambda<0$, it is true by
Eq.~(\ref{HepPlease}) that ${\rm
tr}\big(A_{\vec{q}}\Pi\big)=1+\lambda >1$. Consider repeating this
measurement over and over. In particular, let us tabulate the
probability of getting outcome $\Pi$ for every single trial to the
exclusion of all other outcomes.

The gist of the contradiction is most easily seen by {\it imagining\/}
that Eq.~(\ref{Howard}) is really a discrete sum:
\begin{equation}
\rho^{(N)}= P(\vec{q})\,A_{\vec{q}}^{\otimes
N}+\sum_{\vec{p}\ne\vec{q}}P(\vec{p})\,A_{\vec{p}}^{\otimes N}\;.
\end{equation}
The probability of $N$ occurrences of the outcome $\Pi$ is thus
\begin{eqnarray}
{\rm tr}\big(\rho^{(N)}\Pi^{\otimes N}\big)
&=&
P(\vec{q})\,{\rm tr}(A_{\vec{q}}^{\otimes N}\Pi^{\otimes N})
+\sum_{\vec{p}\ne\vec{q}}P(\vec{p})\,{\rm tr}(A_{\vec{p}}^{\otimes
N}\Pi^{\otimes N})
\nonumber\\
&=&
P(\vec{q})\,[{\rm tr}(A_{\vec{q}}\Pi)]^N
+\sum_{\vec{p}\ne\vec{q}}P(\vec{p})\,[{\rm tr}(A_{\vec{p}}\Pi)]^N
\nonumber\\
&=&
P(\vec{q})(1+\lambda)^N +\sum_{\vec{p}\ne\vec{q}}P(\vec{p})\,[{\rm
tr}(A_{\vec{p}}\Pi)]^N\;.
\label{Hanna}
\end{eqnarray}
There are no assurances in general that the right-hand term in
Eq.~(\ref{Hanna}) is positive, but if $N$ is an even number it
must be.  It follows that if $P(\vec{q})\ge0$, for sufficiently
large {\it even\/} $N$,
\begin{equation}
{\rm tr}\big(\rho^{(N)}\Pi^{\otimes N}\big)>1\;,
\label{BigBoy}
\end{equation}
contradicting the assumption that it should always be a probability.

All we need to do now is transcribe the argument leading to
Eq.~(\ref{BigBoy}) to the general integral case of
Eq.~(\ref{Howard}). Note that by Eq.~(\ref{Hamish}), the quantity
${\rm tr}\big(A_{\vec{p}}\Pi\big)$ is a (linear) continuous
function of the parameter $\vec{p}$.  Therefore, for any
$\epsilon>0$, there exists a $\delta>0$ such that $\big|{\rm
tr}\big(A_{\vec{p}}\Pi\big)- {\rm
tr}\big(A_{\vec{q}}\Pi\big)\big|\le\epsilon$ whenever
$|\vec{p}-\vec{q}|\le\delta$, i.e., whenever $\vec{p}$ is
contained within an open ball $B_\delta(\vec{q})$ centered at
$\vec{q}$.  Choose $\epsilon<\lambda$, and define
$\overline{B}_\delta$ to be the intersection of
$B_\delta(\vec{q})$ with the probability simplex. For $\vec{p}\in
\overline{B}_\delta$, it follows that
\begin{equation}
{\rm tr}\big(A_{\vec{p}}\Pi\big)\ge 1+\lambda-\epsilon>1\;.
\end{equation}
If we consider an $N$ that is even, $\big[{\rm
tr}\big(A_{\vec{p}}\Pi\big)\big]^N$ is nonnegative in all of
${\cal S}_{d^2}$, and we have that the probability of the outcome
$\Pi^{\otimes N}$ satisfies
\begin{eqnarray}
{\rm tr}\big(\rho^{(N)}\Pi^{\otimes N}\big)
&=&
\int_{{\cal S}_{d^2}} P(\vec{p})\,\big[{\rm
tr}\big(A_{\vec{p}}\Pi\big)\big]^N \,d\vec{p}\;
\nonumber\\
&=&
\int_{{\cal S}_{d^2}-\overline{B}_\delta} P(\vec{p})\,\big[{\rm
tr}\big(A_{\vec{p}}\Pi\big)\big]^N \,d\vec{p}
\; +\, \int_{\overline{B}_\delta} P(\vec{p})\,\big[{\rm
tr}\big(A_{\vec{p}}\Pi\big)\big]^N \,d\vec{p}
\nonumber\\
&\ge&
\int_{\overline{B}_\delta} P(\vec{p})\,\big[{\rm
tr}\big(A_{\vec{p}}\Pi\big)\big]^N \,d\vec{p}
\nonumber\\
&\ge&
(1+\lambda-\epsilon)^N \int_{\overline{B}_\delta}
P(\vec{p})\,d\vec{p}\;.
\label{Homer}
\end{eqnarray}
Unless
\begin{equation}
\int_{\overline{B}_\delta} P(\vec{p})\,d\vec{p}=0\;,
\end{equation}
the lower bound (\ref{Homer}) for the probability of the outcome
$\Pi^{\otimes N}$ becomes arbitrarily large as
$N\rightarrow\infty$.  Thus we conclude that the requirement that
$\rho^{(N)}$ be a proper density operator constrains $P(\vec{p})$
to vanish almost everywhere in $\overline{B}_\delta$ and,
consequently, to vanish almost everywhere that $A_{\vec{p}}$ is
not a physical state.

Using Eq.~(\ref{Hamish}), we can trivially transform the integral
representation (\ref{Howard}) to one directly over the convex set
of density operators ${\cal D}_d$ and be left with the following
statement. Under the sole assumption that the density operator
$\rho^{(N)}$ is exchangeable, there exists a unique probability
density $P(\rho)$ such that
\begin{equation}
\rho^{(N)}=\int_{{\cal D}_d} P(\rho)\, \rho^{\otimes N}\, d\rho\;.
\end{equation}
This concludes the proof of the quantum de Finetti representation
theorem.

\section{Intermezzo} \label{sec-intermezzo}

In classical probability theory, exchangeability characterizes those
situations where the only data relevant for updating a probability
distribution are frequency data, i.e., the numbers $n_j$ in
Eq.~(\ref{eq-repr}) which tell how often the result $j$ occurred. The
quantum de Finetti representation shows that the same is true in
quantum mechanics:  Frequency data (with respect to a sufficiently
robust measurement) are sufficient for updating an exchangeable state
to the point where nothing more can be learned from sequential
measurements; that is, one obtains a convergence of the
form~(\ref{HannibalLecter}), so that ultimately any further
measurements on the individual systems are statistically independent.

Beyond the aesthetic point of showing the consistency of the Bayesian
conception of quantum states, we also believe the technical methods
exhibited here will be of interest in the practical arena. Recently
there has been a large literature on which classes of measurements
have various advantages for tomographic
purposes~\cite{QuorumLump,QuorumOld}.  To our knowledge, the work in
Ref.~\cite{Caves2002b} reiterated here was the first to consider
tomographic reconstruction based upon minimal informationally
complete POVMs. One can imagine several advantages to this approach
via the fact that such POVMs with rank-one elements are automatically
extreme points in the convex set of all
measurements~\cite{Fujiwara1998}.

Furthermore, the classical de Finetti theorem is only the beginning
with respect to general questions in classical statistics to do with
exchangeability and generalizations of the concept~\cite{Aldous1985}.
One should expect no less of quantum exchangeability studies.  In
particular, we are thinking of such issues as representation theorems
for finitely exchangeable
distributions~\cite{Jaynes1986,Diaconis1977}. Just as our method for
proving the quantum de Finetti theorem was able to rely heavily on
the classical theorem, so one might expect similar benefits from the
classical results in the case of quantum finite
exchangeability---although, there will certainly be new aspects to
the quantum case due to the possibility of entanglement in finite
exchangeable states~\cite{Emch2002}.  Finally, a practical
application of such representation theorems could be their potential
to contribute to the solution of some outstanding problems in
constructing security proofs for various quantum key distribution
schemes~\cite{Lo2000,Tamaki2003}.

\section{Subjectivity of Quantum Operations} \label{Emma}

At this point we turn our attention to quantum operations.  So far,
we have made much to-do of the subjectivity of quantum states.  But
what of quantum operations?  Should this structural element within
quantum theory be considered of the nature of a fact---like the
outcome of a quantum measurement---or, like the quantum state, should
it be recognized as a statement of an agent's belief or
expectation~\cite{FuchsWHAT}?

The usual presentation of the concept of a quantum operation is that
it is the most general quantum-state evolution allowed by the laws of
quantum mechanics~\cite{Kraus1983,Nielsen2000}.  Upon the process of
measurement of some POVM $\{ E_\alpha\}$, depending upon the
particular details of a measurement interaction and the outcome
$\alpha$, an initial quantum state $\rho$ can change to any new state
of the form
\be
\Phi_\alpha(\rho)=\frac{1}{\,\tr\rho E_\alpha}\sum_i A_{\alpha i}
\rho A_{\alpha i}^\dagger\;,
\ee
where $A_{\alpha i}$ can be any operators whatsoever, as long as they
satisfy
\be
E_\alpha = \sum_i A_{\alpha i}^\dagger A_{\alpha i}\;.
\ee
In the case where the POVM is the completely uninformative
one---i.e., the POVM $\{ I\}$ comprised solely of the identity
operator---one recovers the general expression for all possible
quantum mechanical time evolutions in terms of trace-preserving
completely positive maps:
\be
\rho \; \longrightarrow\; \Phi(\rho)=\frac{1}{\tr\rho
E_\alpha}\sum_\alpha A_{\alpha} \rho A_{\alpha}^\dagger\;,
\ee
where the $A_\alpha$ are any operators that satisfy
\be
I = \sum_\alpha A_{\alpha}^\dagger A_{\alpha}\;.
\ee

Let us first focus on the latter type of evolution. A quantum
Bayesian should become suspicious that it contains a subjective
component because of another representation theorem of
Kraus~\cite{Kraus1983}. For any trace preserving completely positive
map $\Phi$ on a system, one can always imagine an ancillary system
$A$, a quantum state $\sigma$ for that ancillary system, and unitary
interaction $U$ between the system and the ancilla, such that
\be
\Phi(\rho)=\tr_A \!\left( U(\rho\otimes\sigma)U^\dagger \right)\,,
\label{RainyDay}
\ee
where $\tr_A$ represents a partial trace over the ancilla's Hilbert
space.  The Bayesian should ask, ``Whose state of belief is
$\sigma$?''

But worse than that, the representation in Eq.~(\ref{RainyDay}) is
not unique.  There can well be distinct density operators $\tau$ for
the ancilla and distinct unitary interactions $V$ such that
\be
\Phi(\rho)=\tr_A \!\left( V(\rho\otimes\tau)V^\dagger \right)\,.
\label{RainyDayReprise}
\ee
What is to be made of this?  If one accepts that a quantum state is a
subjective judgment, then it would seem one should also be compelled
to accept that the map $\Phi$ can be thought of as having (at least)
a subjective component through the quantum state $\sigma$, and that
the subjective judgment even leaks into unitary operation $U$,
nominally describing the interaction.

Without belaboring that particular point, let us now change focus to the
state-change rule associated with the von Neumann collapse postulate.
In that context, the measurement operators
$E_\alpha=|\alpha\rangle\langle\alpha|$ correspond to projectors onto
an orthonormal basis $\{|\alpha\rangle\}$ and
\be
\Phi_\alpha(\rho) = \frac{1}{\,\tr\rho E_\alpha}\, E_\alpha\rho
E_\alpha = |\alpha\rangle\langle\alpha|\;.
\ee
The salient point is that, conditioned on the measurement outcome,
the final quantum state for the system is uniquely determined after
such a measurement.

Note what this implies.  If the quantum operation associated with a
measurement device is an objective fact (i.e., of the same nature as
the measurement outcome), then so too must be the posterior quantum
state $E_\alpha=|\alpha\rangle\langle\alpha|$.  But the whole
foundation of this paper is that quantum states are not objective
facts.  It follows that the quantum operation $\Phi_\alpha$ cannot be
an objective fact either.  Instead, the von Neumann collapse rule
associated with a projective measurement must be as much of a
subjective judgment as a quantum state is in the first place.

The moral:  (At least some) quantum operations are not facts, and
general quantum time evolutions have a subjective
component.\,\footnote{Ref.~\cite{FuchsWHAT}, in fact, tries to argue
for more:  Namely, that {\it all\/} quantum operations are subjective
states of belief, just like all quantum states are subjective states
of belief.  The present work, however, refrains from attempting that
larger task.} But, if so, then what are the ``unknown quantum
operations'' that experimentalists routinely measure in the
laboratory?

\section{Quantum Process Tomography} \label{Katie}

In quantum process
tomography~\cite{Turchette1995b,Chuang1997,Poyatos1997}, an
experimenter lets an incompletely specified device act on a quantum
system prepared in an input state of his choice, and then performs a
measurement (also of his choice) on the output system. This procedure
is repeated many times over, with possibly different input states and
different measurements, in order to accumulate enough statistics to
assign a quantum operation to the device. Here and throughout this
section, by a quantum operation we mean a trace-preserving completely
positive linear map. Quantum process tomography has been demonstrated
experimentally in liquid state nuclear magnetic resonance
\cite{Nielsen1998a,Childs2001}, and recently a number of optical
experiments \cite{Nambu2002,DeMartini-0210,Altepeter-0303} have
implemented entanglement-assisted quantum process tomography. The
latter is a procedure that exploits the fact that quantum process
tomography is equivalent to quantum state tomography in a larger
state space \cite{Leung2001,Leung2002,Duer2001,D'Ariano2001}.

In the usual description of process tomography, it is assumed that
the device performs the same {\it unknown\/} quantum operation $\Phi$
every time it is used, and an experimenter's prior information about
the device is expressed via a probability density $p(\Phi)$ over all
possible operations.  What, however, is the operational meaning of an
unknown quantum operation?  When does the action of a device leave
off from an initial input so that the next input can be sent through?
In particular, what gives the right to suppose that a device does not
have memory or, for instance, does not entangle the successive inputs
passing through it?  These questions boil down to the need to explore
a single issue: What essential assumptions must be made so that
quantum process tomography is a logically coherent notion?

What is called for is a method of posing quantum process tomography
that never requires the invocation of the concept of an unknown
quantum operation. This can be done by focussing upon the action of a
single {\it known\/} quantum operation $\Phi^{(N)}$, which acts upon
$N$ nominal inputs.  In particular, we identify conditions under
which $\Phi^{(N)}$, ($N=1,2,\ldots$), can be represented as
\begin{equation}   \label{eq:prior}
\Phi^{(N)}=\int p(\Phi)\, \Phi^{\otimes N}\,d\Phi \;,
\label{TheEquationWithNoName}
\end{equation}
for some probability density $p(\Phi)$, and where the integration
extends over all single-system quantum operations $\Phi$.  With this
theorem established, the conditions under which an experimenter can
act {\it as if\/} his prior $\Phi^{(N)}$ corresponds to {\em
ignorance\/} of a ``true'' but unknown quantum operation are made
precise.

Our starting point is the closely aligned and similarly motivated de
Finetti representation theorem for quantum states of the previous
sections. Here, we make use of the correspondence between quantum
process tomography and quantum-state tomography mentioned above to
derive a de Finetti representation theorem for sequences of quantum
operations.

\section{The Process-Tomography Theorem}
\label{sectheorem}

In this section and the next, we restrict our attention to devices
for which the input and output have the same Hilbert space dimension,
$D$.  In the following, $\HD$ denotes a $D$-dimensional Hilbert
space, $\HD^{\otimes N}=\HD\otimes\cdots\otimes\HD$ denotes its
$N$-fold tensor product, and ${\cal L}({\cal V})$ denotes the space
of linear operators on a linear space ${\cal V}$. The set of density
operators for a $D$-dimensional quantum system is a convex subset of
${\cal
  L}(\HD)$.

The action of a device on $N$ nominal inputs systems is then
described by a trace-preserving completely positive map
\begin{equation}
\Phi^{(N)} : {\cal L}(\HD^{\otimes N}) \longrightarrow
  {\cal L}(\HD^{\otimes N}) \;,
\end{equation}
which maps the state of the $N$ input systems to the state of the $N$
output systems. We will say, in analogy to the definition of
exchangeability for quantum states, that a quantum operation
$\Phi^{(N)}$ is {\it
  exchangeable\/} if it is a member of an exchangeable sequence of
quantum operations.

To define exchangeability for a sequence of quantum operations in a
natural way, we reduce the properties of symmetry and extendibility
for sequences of operations to the corresponding properties for
sequences of states. In the following, we will use bold letters to
denote vectors of indices, e.g.  ${\bf
  j}=(j_1,\ldots,j_N)$. We will use $\pi$ to denote a permutation of the set
$\{1,\ldots,N\}$, where the cardinality $N$ will depend on the
context. The action of the permutation $\pi$ on the vector ${\bf j}$
is defined by $\pi{\bf j} = (j_{\pi(1)},\ldots,j_{\pi(N)})$.

Any $N$-system density operator $\rho^{(N)}$ can  be expanded in the
form
\begin{equation}
\rho^{(N)}=\sum_{{\bf j},{\bf l}} r^{(N)}_{{\bf j},{\bf l}}
\bigotimes_{i=1}^N |j_i^{Q_i}\rangle\langle l_i^{Q_i}| \equiv
\sum_{{\bf j},{\bf l}} r^{(N)}_{{\bf j},{\bf l}}
|j_1^{Q_1}\rangle\langle l_1^{Q_1}|
  \otimes\cdots\otimes |j_N^{Q_N}\rangle\langle l_N^{Q_N}|  \;,
\end{equation}
where $\{\ket{1^{Q_i}},\ldots,\ket{D^{Q_i}}\}$ denotes an orthonormal
basis for the Hilbert space $\HD$ of the $i$th system, and
$r^{(N)}_{{\bf j},{\bf l}}$ are the matrix elements of $\rho^{(N)}$
in the tensor product basis. We define the action of the permutation
$\pi$ on the state $\rho^{(N)}$ by
\begin{equation}
\pi\rho^{(N)}=\sum_{{\bf j},{\bf l}} r^{(N)}_{\pi{\bf j},\pi{\bf l}}
\bigotimes_{i=1}^N |j_i^{Q_i}\rangle\langle l_i^{Q_i}| =\sum_{{\bf
j},{\bf l}} r^{(N)}_{{\bf j},{\bf l}} \bigotimes_{i=1}^N
|j_{\pi^{-1}(i)}^{Q_i}\rangle\langle l_{\pi^{-1}(i)}^{Q_i}| \;.
\end{equation}

With this notation, we can make the following definition. A sequence
of quantum operations, $\Phi^{(k)} : {\cal L}(\HD^{\otimes k})
\rightarrow {\cal L}(\HD^{\otimes k})$, is called {\it
exchangeable\/} if, for $k=1,2,\ldots$,
\begin{enumerate}
\item $\Phi^{(k)}$ is symmetric, i.e.,
\begin{equation}   \label{eq:symmetric}
\Phi^{(k)}(\rho^{(k)}) = \pi\Big(\Phi^{(k)}(\pi^{-1}\rho^{(k)})\Big)
\end{equation}
for any permutation $\pi$ of the set $\{1,\ldots,k\}$ and for any
density operator $\rho^{(k)} \in {\cal L}(\HD^{\otimes k})$, and
\item
$\Phi^{(k)}$ is extendible, i.e.,
\begin{equation}  \label{eq:extendible}
\Phi^{(k)}(\tr_{k+1}\rho^{(k+1)}) =
\tr_{k+1}\Big(\Phi^{(k+1)}(\rho^{(k+1)})\Big)
\end{equation}
for any state $\rho^{(k+1)}$.
\end{enumerate}
In words, these conditions amount to the following. Condition (1) is
equivalent to the requirement that the quantum operation $\Phi^{(k)}$
commutes with any permutation operator $\pi$ acting on the states
$\rho^{(k)}$:  It does not matter what order we send our systems
through the device; as long as we rearrange them at the end into the
original order, the resulting evolution will be the same. Condition
(2) says that it does not matter if we consider a larger map
$\Phi^{(N+1)}$ acting on a larger collection of systems (possibly
entangled), or a smaller $\Phi^{(N)}$ on some subset of those
systems:  The upshot of the evolution will be the same for the
relevant systems.

We are now in a position to formulate the de Finetti representation
theorem for quantum operations. A quantum operation $\Phi^{(N)} :
{\cal L}(\HD^{\otimes N}) \rightarrow {\cal L}(\HD^{\otimes N})$ is
an element of an exchangeable sequence if and only if it can be
written in the form
\begin{equation}
\Phi^{(N)}=\int p(\Phi)\, \Phi^{\otimes N}\, d\Phi\, \quad \mbox{for
all $N$},
\end{equation}
where the integral ranges over all single-shot quantum operations
$\Phi:{\cal L}(\HD)\rightarrow{\cal L}(\HD)$, $d\Phi$ is a suitable
measure on the space of quantum operations, and the probability
density $p(\Phi)\ge0$ is unique. The tensor product $\Phi^{\otimes
N}$ is defined by $\Phi^{\otimes N}(\rho_1\otimes\cdots\otimes\rho_N)
=\Phi(\rho_1)\otimes\cdots\otimes\Phi(\rho_N)$ for all
$\rho_1,\ldots,\rho_N$ and by linear extension for arbitrary
arguments.

Just as with the original quantum de Finetti theorem
\cite{Hudson1976,Caves2002b}, this result allows a certain latitude
in how quantum process tomography can be described.  One is free to
use the language of an unknown quantum operation if the condition of
exchangeability is met by one's prior $\Phi^{(N)}$ but it is not
required:  For the quantum Bayesian in particular, the {\it known\/}
quantum operation $\Phi^{(N)}$ is the only meaningful quantum
operation in the problem.

\section{Proof of the Process-Tomography Theorem}  \label{secproof}

Let $\Phi^{(N)}$, $N=1,2,\ldots$, be an exchangeable sequence of
quantum operations. $\Phi^{(N)}$ can be characterized in terms of its
action on the elements of a basis of ${\cal L}(\HD^{\otimes N})$ as
follows.
\begin{equation}   \label{eq:sljmk}
\Phi^{(N)} \Big( \bigotimes_{i=1}^N \ket{j_i^{Q_i}}\bra{k_i^{Q_i}}
\Big) =\sum_{{\bf l},{\bf m}}
 S^{(N)}_{{\bf l},{\bf j},{\bf m},{\bf k}}
       \bigotimes_{i=1}^N \ket{l_i^{Q_i}}\bra{m_i^{Q_i}} \;.
\end{equation}
The coefficients $S^{(N)}_{{\bf l},{\bf j},{\bf m},{\bf k}}$ specify
$\Phi^{(N)}$ uniquely.  It follows from a construction due to
Choi~\cite{Choi1975} that the $S^{(N)}_{{\bf l},{\bf j},{\bf m},{\bf
    k}}$ can be regarded as the matrix elements of a density operator on
$D^{2N}$-dimensional Hilbert space $\H_{D^2}^{\otimes N}$. This can
be seen as follows.  Let
\begin{equation}
\ket\Psi = {1\over\sqrt D} \sum_{k=1}^D \ket{k^{R_i}}\ket{k^{Q_i}}
   \in \HD\otimes\HD= \H_{D^2}
\end{equation}
be a maximally entangled state in $\H_{D^2}$, where the
$\ket{k^{R_i}}$ ($k=1,\ldots,D$) form orthonormal bases for the
ancillary systems labelled $R_i$ ($i=1,\ldots,N$). The corresponding
density operator is
\begin{equation}
\ket\Psi\bra\Psi = {1\over D} \sum_{j,k} \ket{j^{R_i}}\bra{k^{R_i}}
   \otimes\ket{j^{Q_i}}\bra{k^{Q_i}} \in {\cal
  L}(\H_{D^2}) \;.
\end{equation}
Similarly, we define a map, $J$, from the set of quantum operations
on $\HD^{\otimes N}$ to the set of density operators on
$\H_{D^2}^{\otimes N}$ by
\begin{eqnarray}   \label{eq:Jdef}
J ( \Phi^{(N)} ) &\equiv& \Big(I^{(N)} \otimes \Phi^{(N)}\Big)
   \Big( (\ket\Psi\bra\Psi)^{\otimes N} \Big)  \cr
&=&
{1\over D^N}\Big(I^{(N)} \otimes \Phi^{(N)}\Big)
    \Big( \sum_{{\bf j},{\bf k}} \bigotimes_{i=1}^N
(\ket{j_i^{R_i}}\bra{k_i^{R_i}} \otimes
\ket{j_i^{Q_i}}\bra{k_i^{Q_i}}) \Big)
  \cr
&=&
{1\over D^N} \sum_{{\bf l},{\bf j},{\bf m},{\bf k}}
 S^{(N)}_{{\bf l},{\bf j},{\bf m},{\bf k}} \bigotimes_{i=1}^N
(\ket{j_i^{R_i}}\bra{k_i^{R_i}} \otimes
\ket{l_i^{Q_i}}\bra{m_i^{Q_i}}) \;.
\end{eqnarray}
In this definition, $I^{(N)}$ denotes the identity operation acting
on the ancillary systems $R_1,\ldots,R_N$.  The map $J$ is injective,
i.e. $J(\Phi_1^{(N)})=J(\Phi_2^{(N)})$ if and only if
$\Phi_1^{(N)}=\Phi_2^{(N)}$.

The first stage of the proof of the de Finetti theorem for operations
is to show that the density operators $J(\Phi^{(N)})$,
$N=1,2,\ldots$, form an exchangeable sequence when regarded as
$N$-system states, with $R_i$ and $Q_i$ jointly forming the $i$th
system.  To do this, we first show that $J(\Phi^{(N)})$ is symmetric,
i.e., invariant under an arbitrary permutation $\pi$ of the $N$
systems.

Note that since the density operators $\rho^{(N)}$ actually span the
whole vector space ${\cal L}(\HD^{\otimes N})$, enforcing Definition
1 above amounts to identifying the linear maps on the left- and
right-hand sides of Eqs.~(\ref{eq:symmetric}) and
(\ref{eq:extendible}).  I.e.,
\begin{equation}
\Phi^{(k)} = \pi\circ\Phi^{(k)}\circ\pi^{-1}
\end{equation}
and
\begin{equation}
\Phi^{(k)}\circ\tr_{k+1} = \tr_{k+1}\circ\Phi^{(k+1)}
\end{equation}
Thus in much that we do it suffices to consider the action of these
maps on an arbitrary basis state $E^{(N)}=\bigotimes_{i=1}^N
\ket{j_i^{Q_i}}\bra{k_i^{Q_i}}$ for arbitrary ${\bf j}$ and ${\bf
k}$. In particular,
\begin{eqnarray}   \label{eq:pisljmk}
\pi\Big(\Phi^{(N)}(\pi^{-1}E^{(N)})\Big)
&=&
\pi\Big(\Phi^{(N)}\Big(
 \bigotimes_{i=1}^N \ket{j_{\pi(i)}^{Q_i}}\bra{k_{\pi(i)}^{Q_i}}
 \Big)\Big)  \cr
&=&
\pi\sum_{{\bf l},{\bf m}}
 S^{(N)}_{{\bf l},\pi{\bf j},{\bf m},\pi{\bf k}}
       \bigotimes_{i=1}^N \ket{l_i^{Q_i}}\bra{m_i^{Q_i}} \cr
&=&
\sum_{{\bf l},{\bf m}}
 S^{(N)}_{\pi{\bf l},\pi{\bf j},\pi{\bf m},\pi{\bf k}}
       \bigotimes_{i=1}^N \ket{l_i^{Q_i}}\bra{m_i^{Q_i}} \;.
\end{eqnarray}
Assuming Eq.~(\ref{eq:symmetric}), i.e., symmetry of $\Phi^{(N)}$,
for all ${\bf j}$ and ${\bf k}$, it follows that
\begin{equation}
S^{(N)}_{\pi{\bf l},\pi{\bf j},\pi{\bf m},\pi{\bf k}} =
S^{(N)}_{{\bf l},{\bf j},{\bf m},{\bf k}}
\end{equation}
for all ${\bf l},{\bf j},{\bf m},{\bf k}$, which, using
Eq.~(\ref{eq:Jdef}), implies that
\begin{equation}
\pi(J ( \Phi^{(N)} )) =  J ( \Phi^{(N)} ) \;,
\end{equation}
i.e., symmetry of $J(\Phi^{(N)})$.

To prove extendibility of $J(\Phi^{(N)})$, we introduce the following
notation for partial traces: we denote by $\tr_{N+1}^R$ the partial
trace over the subsystem $R_{N+1}$, and by $\tr_{N+1}^Q$ the partial
trace over the subsystem $Q_{N+1}$. In this notation, we need to show
that $\tr_{N+1}^R\tr_{N+1}^Q J(\Phi^{(N+1)})=J(\Phi^{(N)})$. Using
Eqs.~(\ref{eq:extendible}) and~(\ref{eq:Jdef}),
\begin{eqnarray}
&& \tr_{N+1}^R\tr_{N+1}^Q J(\Phi^{(N+1)}) \cr
&& =
\tr_{N+1}^R\tr_{N+1}^Q{1\over D^{N+1}}\Big(I^{(N+1)} \otimes
\Phi^{(N+1)}\Big)
    \Big( \sum_{{\bf j},j_{N+1},{\bf k},k_{N+1}} \bigotimes_{i=1}^{N+1}
(\ket{j_i^{R_i}}\bra{k_i^{R_i}} \otimes
\ket{j_i^{Q_i}}\bra{k_i^{Q_i}}) \Big)
\cr
&& =
\tr_{N+1}^Q{1\over D^{N+1}}\Big(I^{(N)} \otimes \Phi^{(N+1)}\Big)
    \Big( \sum_{{\bf j},{\bf k},k_{N+1}} \bigotimes_{i=1}^{N}
(\ket{j_i^{R_i}}\bra{k_i^{R_i}} \otimes
\ket{j_i^{Q_i}}\bra{k_i^{Q_i}})
 \otimes \ket{k_{N+1}^{Q_{N+1}}}\bra{k_{N+1}^{Q_{N+1}}}  \Big)
\cr
&& =
{1\over D^{N+1}} \sum_{{\bf j},{\bf k},k_{N+1}} \Big(
\bigotimes_{i=1}^{N} (\ket{j_i^{R_i}}\bra{k_i^{R_i}}\Big) \otimes
\tr_{N+1}^Q\Phi^{(N+1)}\Big(
\bigotimes_{l=1}^{N}\ket{j_l^{Q_i}}\bra{k_l^{Q_i}}
 \otimes \ket{k_{N+1}^{Q_{N+1}}}\bra{k_{N+1}^{Q_{N+1}}}\Big)
\cr
&& =
{1\over D^{N+1}} \sum_{{\bf j},{\bf k},k_{N+1}} \Big(
\bigotimes_{i=1}^{N} (\ket{j_i^{R_i}}\bra{k_i^{R_i}}\Big) \otimes
\Phi^{(N)}\Big( \bigotimes_{l=1}^{N}\ket{j_l^{Q_i}}\bra{k_l^{Q_i}}
 \Big)
\cr
&& =
{1\over D^{N+1}}\Big(I^{(N)} \otimes \Phi^{(N)}\Big)
    \Big(  \sum_{{\bf j},{\bf k},k_{N+1}}
  \bigotimes_{i=1}^{N}
(\ket{j_i^{R_i}}\bra{k_i^{R_i}} \otimes
\ket{j_i^{Q_i}}\bra{k_i^{Q_i}}) \Big)
\cr
&& =
{1\over D^N}\Big(I^{(N)} \otimes \Phi^{(N)}\Big)
    \Big(  \sum_{{\bf j},{\bf k}} \bigotimes_{i=1}^{N}
(\ket{j_i^{R_i}}\bra{k_i^{R_i}} \otimes
\ket{j_i^{Q_i}}\bra{k_i^{Q_i}}) \Big)
\cr
&& =
J(\Phi^{(N)}) \;.
\end{eqnarray}

We have thus shown that $J(\Phi^{(N)})$, $N=1,2,\ldots$, form an
exchangeable sequence. According to the quantum de Finetti theorem
for density operators, we can write
\begin{equation}   \label{eq:JdeFinetti}
J(\Phi^{(N)})  = \int  p(\rho)\, \rho^{\otimes N}\,d\rho\;,
\end{equation}
where $p(\rho)\ge0$ is unique, and $\int d\rho\; p(\rho)=1$. With the
parameterization
\begin{equation}   \label{eq:rhoparam}
\rho = {1\over D} \sum_{l,j,m,k} S^{(1)}_{l,j,m,k}
\ket{j^{R}}\bra{k^{R}}
   \otimes\ket{l^{Q}}\bra{m^{Q}} \;,
\end{equation}
Eq.~(\ref{eq:JdeFinetti}) takes the form
\begin{eqnarray}    \label{eq:Jparametrized}
J(\Phi^{(N)})
&=& {1\over D^N}\int_{{\cal D}} dS \; p(S) \;
\Big(\sum_{l,j,m,k} S^{(1)}_{l,j,m,k} \ket{j^{R}}\bra{k^{R}}
   \otimes\ket{l^{Q}}\bra{m^{Q}}  \Big)^{\otimes N} \cr
&=& {1\over D^N}\int_{{\cal D}} dS \; p(S) \;
\bigotimes_{i=1}^N \sum_{l_i,j_i,m_i,k_i} S^{(1)}_{l_i,j_i,m_i,k_i}
\ket{j_i^{R_i}}\bra{k_i^{R_i}}
   \otimes\ket{l_i^{Q_i}}\bra{m_i^{Q_i}} \cr
&=& {1\over D^N} \sum_{{\bf l},{\bf j},{\bf m},{\bf k}}
   \int_{{\cal D}} dS \; p(S) \;
\bigotimes_{i=1}^N S^{(1)}_{l_i,j_i,m_i,k_i}
  \ket{j_i^{R_i}}\bra{k_i^{R_i}}  \otimes\ket{l_i^{Q_i}}\bra{m_i^{Q_i}} \;,
\end{eqnarray}
where the integration variable is a vector with $D^4$ components,
$S=(S^{(1)}_{1,1,1,1},\ldots,S^{(1)}_{D,D,D,D})$, and where the
integration domain, ${\cal D}$, is the set of all $S$ that represent
matrix elements of a density operator. The function $p(S)$ is unique,
$p(S)\ge0$, and $\displaystyle{\int_{{\cal D}} dS\; p(S)=1}$. Notice
the slight abuse of notation in the first line of
Eq.~(\ref{eq:Jparametrized}), where the superscripts $R$ and $Q$
label the entire sequences of systems $R_1,\ldots,R_N$ and
$Q_1,\ldots,Q_N$, respectively.

Comparing Eq.~(\ref{eq:Jparametrized}) with Eq.~(\ref{eq:Jdef}), we
can express the coefficients $S^{(N)}_{{\bf l},{\bf j},{\bf m},{\bf
k}}$ specifying the quantum operation $\Phi^{(N)}$ [see
Eq.~(\ref{eq:sljmk})] in terms of the integral above:
\begin{equation}
S^{(N)}_{{\bf l},{\bf j},{\bf m},{\bf k}} = \int_{{\cal D}} dS \;
p(S) \;
   \prod_{i=1}^N S^{(1)}_{l_i,j_i,m_i,k_i} \;.
\end{equation}
Hence, for any ${\bf j}$ and ${\bf k}$,
\begin{eqnarray}
\Phi^{(N)} \Big( \bigotimes_{i=1}^N \ket{j_i^{Q_i}}\bra{k_i^{Q_i}}
\Big)
&=&  \sum_{{\bf l},{\bf m}} \int_{{\cal D}} dS \; p(S) \;
   \Big(\prod_{i=1}^N S^{(1)}_{l_i,j_i,m_i,k_i} \Big)
  \bigotimes_{i=1}^N \ket{l_i^{Q_i}}\bra{m_i^{Q_i}}    \cr
&=&  \int_{{\cal D}} dS \; p(S) \;
   \bigotimes_{i=1}^N
  \sum_{l_i,m_i} S^{(1)}_{l_i,j_i,m_i,k_i}\ket{l_i^{Q_i}}\bra{m_i^{Q_i}} \;.
\end{eqnarray}
The $D^4$ coefficients, $S^{(1)}_{l,j,m,k}$,  of the vector $S$
define a single-system map, $\Phi_S$, via
\begin{equation}
\Phi_S(\ket{j^{Q}}\bra{k^{Q}})  \equiv
  \sum_{l,m} S^{(1)}_{l,j,m,k}\ket{l^{Q}}\bra{m^{Q}} \;\;\;(j,k=1,\ldots,D)\;.
\end{equation}
Hence
\begin{eqnarray}  \label{eq:predefinetti}
\Phi^{(N)} \Big( \bigotimes_{i=1}^N \ket{j_i^{Q_i}}\bra{k_i^{Q_i}}
\Big)
&=& \int_{{\cal D}} dS \; p(S) \;
   \bigotimes_{i=1}^N
\Phi_S\Big(\ket{j_i^{Q_i}}\bra{k_i^{Q_i}}\Big) \cr
&=& \int_{{\cal D}} dS \; p(S) \;      \Phi_S^{\otimes N}
 \Big( \bigotimes_{i=1}^N \ket{j_i^{Q_i}}\bra{k_i^{Q_i}}\Big) \;.
\end{eqnarray}
Since this equality holds for arbitrary ${\bf j}$ and ${\bf k}$, it
implies the representation
\begin{equation}  \label{eq:finaldefinetti}
\Phi^{(N)}  = \int_{{\cal D}} dS \; p(S) \;   \Phi_S^{\otimes N} \;.
\end{equation}
For all $S\in {\cal D}$, the map $\Phi_S$ is completely positive.
This can be seen by considering
$$J(\Phi_S) = (I\otimes\Phi_S)\big(\ket\Psi\bra\Psi\big) = {1\over D}
\sum_{l,j,m,k} S^{(1)}_{l,j,m,k} \ket{j^{R}}\bra{k^{R}}
\otimes\ket{l^{Q}}\bra{m^{Q}} \;,$$ which,  by definition of ${\cal
D}$,  is a density operator and therefore positive. It follows from a
theorem by Choi \cite{Choi1975} that $\Phi_S$ is completely positive.

To complete the proof, we will now show that $p(S)=0$ almost
everywhere unless $\Phi_S$ is trace-preserving, i.e., a quantum
operation.  More precisely, we show that if $U\in{\cal D}$ is such
that $\Phi_U$ is not trace-preserving, then there exists an open ball
$B$ containing $U$ such that $p(S)=0$ in $B\cap{\cal D}$.

The essence of the argument can be most easily explained in the
special case that the integral~(\ref{TheEquationWithNoName}) takes
the form of a sum,
\begin{equation}    \label{eq:assume}
\Phi^{(N)} = { \sum_i p_i \; \Phi_i^{\otimes N}} \;,
\end{equation}
where $p_i>0$. It follows that
\begin{equation}   \label{eq:follows}
1 = \sum_i p_i \big(\tr[\Phi_i(\rho)]\big)^N
\end{equation}
for all single-system density operators $\rho$. Now assume that the
sum extends over some non-trace-preserving operation, which we take
to be $\Phi_1$ without loss of generality. This means that
\begin{equation}
\tr[\Phi_1(\rho)]\ne1
\end{equation}
for some single-system density operator $\rho$. Now either
$\tr[\Phi_1(\rho)]<1$, in which case normalization of
$\Phi^{(1)}(\rho)$ implies that $\tr[\Phi_k(\rho)]>1$ for some
$k\ne1$, or $\tr[\Phi_1(\rho)]>1$, in which case we set $k=1$. In
both cases
\begin{equation}
p_k \big(\tr[\Phi_k(\rho)]\big)^N \rightarrow \infty \;,
\end{equation}
which contradicts Eq.~(\ref{eq:follows}). We have thus shown that the
sum~(\ref{eq:assume}) extends only over trace-preserving operations.

Now let us return to the general case where $\Phi^{(N)}$ is
represented by an integral. For $\delta>0$ and $U\in{\cal D}$, we
define $B_\delta(U)$ to be the set of all $S$ such that
$|S-U|<\delta$, i.e.,  $B_\delta(U)$ is the open ball of radius
$\delta$ centered at $U$. Furthermore, we define $\bar
B_\delta(U)=B_\delta(U)\cap{\cal D}$.

Let $U\in{\cal D}$ be such that $\Phi_U$ is not trace-preserving,
i.e., there exists a density operator $\rho$ for which
$\tr[\Phi_U(\rho)]\ne1$. We distinguish two cases.

\vspace{3mm}\noindent{\bf Case (i)}:
$\tr[\Phi_U(\rho)]=1+\epsilon$, where $\epsilon>0$. Since
$\tr[\Phi_S(\rho)]$ is a linear and therefore continuous function of
the vector $S$, there exists $\delta>0$ such that
\begin{equation}
\Big|\tr[\Phi_S(\rho)]-\tr[\Phi_U(\rho)]\Big|<\epsilon/2
\end{equation}
whenever $S\in B_\delta(U)$. For $S\in\bar B_\delta(U)$,
\begin{equation}
\tr[\Phi_S(\rho)]>1+\epsilon-\epsilon/2=1+\epsilon/2 \;.
\end{equation}
Therefore
\begin{eqnarray}
\tr\big[\Phi^{(N)}(\rho^{\otimes N})\big]
&=&\tr\left[\int_{{\cal D}} dS \; p(S) \; \Phi_S^{\otimes N}(\rho^{\otimes N})\right] \cr
&=&\int_{{\cal D}} dS \; p(S) \; \left(\tr[\Phi_S(\rho)]\right)^N \cr
&=&\int_{{\cal D}\backslash \bar B_\delta(U)}
          dS \; p(S) \; \left(\tr[\Phi_S(\rho)]\right)^N
   + \int_{\bar B_\delta(U)}  dS \; p(S) \; \left(\tr[\Phi_S(\rho)]\right)^N \cr
&\ge& \int_{\bar B_\delta(U)}  dS \; p(S) \; \left(\tr[\Phi_S(\rho)]\right)^N\cr
&>&   (1+\epsilon/2)^N \int_{\bar B_\delta(U)}  dS \; p(S) \;.
\end{eqnarray}
Unless $\displaystyle{\int_{\bar B_\delta(U)} dS \; p(S) =0}$, there
exists $N$ such that $\tr\big[\Phi^{(N)}(\rho^{\otimes N})\big]>1$,
which contradicts the assumption that $\Phi^{(N)}$ is
trace-preserving. Hence $p(S)=0$ almost everywhere in $\bar
B_\delta(U)$.

\vspace{3mm}\noindent{\bf Case (ii)}:
$\tr[\Phi_U(\rho)]=1-\epsilon$, where $0<\epsilon\le1$. Because of
continuity, there exists $\delta>0$ such that
\begin{equation}
\Big|\tr[\Phi_S(\rho)]-\tr[\Phi_U(\rho)]\Big|<\epsilon/2
\end{equation}
whenever $S\in B_\delta(U)$. Hence, for $S\in\bar B_\delta(U)$,
\begin{equation}
\tr[\Phi_S(\rho)]<1-\epsilon+\epsilon/2=1-\epsilon/2 \;.
\end{equation}
Now assume that $\displaystyle{\int_{\bar B_\delta(U)} dS \; p(S)
=\eta>0}$. Then, letting $N=1$,
\begin{eqnarray}
1 = \tr\big[\Phi^{(1)}(\rho)\big]
&=&\tr\left[\int_{{\cal D}} dS \; p(S) \; \Phi_S(\rho)\right] \cr
&=&\int_{{\cal D}\backslash \bar B_\delta(U)}
          dS \; p(S) \; \tr[\Phi_S(\rho)]
   + \int_{\bar B_\delta(U)}  dS \; p(S) \; \tr[\Phi_S(\rho)] \cr
&<&
\int_{{\cal D}\backslash \bar B_\delta(U)}
    dS \; p(S) \; \tr[\Phi_S(\rho)] + \eta(1-\epsilon/2) \;,
\end{eqnarray}
which implies that
\begin{equation}
\int_{{\cal D}\backslash \bar B_\delta(U)}
dS\;p(S)\;\tr[\Phi_S(\rho)] > 1-\eta + \eta\epsilon/2 > 1-\eta \;.
\end{equation}
Since
\begin{equation}
\int_{{\cal D}\backslash \bar B_\delta(U)} dS\;p(S)=1-\eta \;,
\end{equation}
it follows that there exist $\zeta>0$ and a point $V\in {\cal
D}\backslash \bar B_\delta(U)$ such that $\tr[\Phi_V(\rho)]>1$ and
\begin{equation}
\int_{\bar B_\xi(V)} dS\;p(S) > 0 \;\;\mbox{ for all } \xi\le\zeta
\;.
\end{equation}
We are thus back to case (i) above. Repeating the argument of case
(i) one can show that this contradicts the assumption that
$\Phi^{(N)}$ is trace preserving for large $N$. It follows that
$\eta=0$, i.e., $p(S)=0$ almost everywhere in $\bar B_\delta(U)$.
This concludes the proof of the de Finetti theorem for quantum
operations.

What we have proven here is a representation theorem.  It shows us
when an experimenter is warranted to think of his (prior) {\it
known\/} quantum operation assignment as built out of a lack of
knowledge of a ``true'' but {\it unknown\/} one.  In that way, the
theorem has the same kind of attraction as the previous de Finetti
theorem for quantum states.

In particular for a Bayesian interpretation of quantum mechanics, it
may be a necessary ingredient for its very consistency.  In
Refs.~\cite{FuchsWHAT,Fuchs2002}, it has been argued strenuously that
quantum operations should be considered of essentially the same
physical meaning and status as quantum states themselves:  They are
Bayesian expressions of an experimenter's judgment.  This could be
captured in the slogan ``a quantum operation is really a quantum
state in disguise.'' In other words, the Choi representation theorem
\cite{Choi1975} is not just a mathematical nicety, but is instead of
deep physical significance.\,\footnote{There have been a few pieces
of recent technical work that may be useful for shoring up this idea.
On the contingency, see Ref.~\cite{SaltBin}.}

Therefore, just as an unknown quantum state is an oxymoron in a
Bayesian interpretation of quantum mechanics, so should be an unknown
quantum operation.  In the case of quantum states, the conundrum is
solved by the existence of a de Finetti theorem for quantum
tomography.  In this section we have shown that the conundrum in
quantum process tomography can be solved in the same way.

\section{Concluding Remarks}  \label{Kiki}

Is there something in nature even when there are no observers or
agents about?  At the practical level, it would seem hard to deny
this, and neither of the authors wish to be viewed as doing so. The
world persists without the observer---there is no doubt in either of
our minds about that.  But then, does that require that two of the
most celebrated elements (namely, quantum states and operations) in
quantum theory---our best, most all-encompassing scientific theory to
date---must be viewed as objective, agent-independent constructs?
There is no reason to do so, we say.  In fact, we think there is
everything to be gained from carefully delineating which part of the
structure of quantum theory is about the world and which part is
about the agent's interface with the world.

{}From this perspective, much---{\it but not all}---of quantum
mechanics is about disciplined uncertainty accounting, just as is
Bayesian probability theory in general.  Bernardo and Smith
\cite{Bernardo1994} write this of Bayesian theory,
\bq
What is the nature and scope of Bayesian Statistics ... ?

Bayesian Statistics offers a rationalist theory of personalistic
beliefs in contexts of uncertainty, with the central aim of
characterising how an individual should act in order to avoid
certain kinds of undesirable behavioural inconsistencies.  The
theory establishes that expected utility maximization provides
the basis for rational decision making and that Bayes' theorem
provides the key to the ways in which beliefs should fit together
in the light of changing evidence.  The goal, in effect, is to
establish rules and procedures for individuals concerned with
disciplined uncertainty accounting.  The theory is not descriptive,
in the sense of claiming to model actual behaviour.  Rather, it is
prescriptive, in the sense of saying ``if you wish to avoid the
possibility of these undesirable consequences you must act in the
following way.
\eq
In fact, one might go further and say of quantum theory, that in
those cases where it is not just Bayesian probability theory full
stop, it is a theory of stimulation and
response~\cite{FuchsWHAT,FuchsPaulian}. The agent, through the
process of quantum measurement stimulates the world external to
himself.  The world, in return, stimulates a response in the agent
that is quantified by a change in his beliefs---i.e., by a change
from a prior to a posterior quantum state.  Somewhere in the
structure of those belief changes lies quantum theory's most direct
statement about what we believe of the world as it is without agents.

The present effort, showing how a Bayesian account of quantum states
and operations is fully consistent with the laboratory practices of
quantum-state and process tomography, is a necessary exercise along
the way to pinpointing that direct statement.

\acknowledgments

We thank Carl Caves and David Mermin for the extensive discussions
that brought part of this paper about, and Marcos P\'erez-Su\'arez
for a careful reading of the manuscript. CAF thanks Science
Foundation Ireland for the support of an E.~T.~S. Walton Visitor
Award.

\end{document}